\documentclass[
 reprint,
 amsmath,
 amssymb,
 aps,
]{revtex4-1}

\usepackage[pdftex]{graphicx}
\usepackage{wasysym}
\usepackage{amsfonts}
\usepackage{ifsym}

\newcommand{\nn}{\nonumber}

\newcommand{\COMMENT}[1]{}

\newcommand{\neqa}{\nonumber\end{eqnarray}}
\newcommand{\la}[1]{\label{#1}}

\def\tr{{\rm tr~}}

\newcommand{\<}{{\langle}}
\renewcommand{\>}{{\rangle}}

\newcommand{\re}{\relax{\rm I\kern-.18em R}}

\def\su2{{SU(2)}}

 \def\mm{\kappa}
\def\V{{\mathfrak{h}}}
\def\N{{N}}
\def\kap{{\kappa}}

\def\ky{{\chi}}
\def\S{\mathcal{S}}
\def\gam{\gamma}

\def\[{\left[}
\def\]{\right]}
\def\Tr{\text{Tr\,}}

\def\({\left(}
\def\){\right)}
\def\[{\left[}
\def\]{\right]}

\def\<{\langle}
\def\>{\rangle}

\def\i2{\frac{i}{2}}

\def\2F1{\,_2{\rm F}_1}

\usepackage[usenames,dvipsnames]{xcolor}
\usepackage{color}
\usepackage{graphicx}
\usepackage{dcolumn}
\usepackage{bm}

\newcommand{\beq}{\begin{equation}}
\newcommand{\eeq}{\end{equation}}
\newcommand{\beqq}{\begin{equation*}}
\newcommand{\eeqq}{\end{equation*}}
\newcommand\beqa{\begin{eqnarray}}
\newcommand\eeqa{\end{eqnarray}}
\newcommand\beqaa{\begin{eqnarray*}}
\newcommand\eeqaa{\end{eqnarray*}}
\newcommand\bea{\begin{array}}
\newcommand\eea{\end{array}}

\begin{document}

\title{Structure Constants and Integrable Bootstrap in Planar $\mathcal{N}=4$ SYM Theory}

\author{Benjamin Basso$^{{\mathcal{O}_1}{\mathcal{O}_2}}$, Shota Komatsu$^{{\mathcal{O}_2}{\mathcal{O}_3}}$ and Pedro Vieira$^{{\mathcal{O}_2}{\mathcal{O}_3}}$}

\affiliation{
\vspace{5mm}
$^{\mathcal{O}_1}$Laboratoire de Physique Th\'eorique, \'Ecole Normale Sup\'erieure, Paris 75005, France\\
$^{\mathcal{O}_2}$ICTP South American Institute for Fundamental Research, IFT-UNESP, S\~ao Paulo, SP Brazil 01440-070 \\
$^{\mathcal{O}_3}$Perimeter Institute for Theoretical Physics, Waterloo, Ontario N2L 2Y5, Canada
}
\begin{abstract}
We introduce a non-perturbative framework for computing structure constants of single-trace operators in the ${\cal N}=4$ SYM theory at large $N$. Our approach features new vertices, with \textit{hexagonal} shape, that can be patched together into three- and possibly higher-point correlators. These newborn hexagons are more elementary and easier to deal with than the three-point functions. Moreover, they can be entirely constructed using integrability, by means of a suitable bootstrap program. In this letter, we present our main results and conjectures for these vertices, and match their predictions for the three-point functions with both weak and strong coupling data available in the literature.
\end{abstract}

\maketitle

\section{Introduction} 

Integrability has emerged as a powerful tool for analyzing maximally supersymmetric gauge theories and their string theory duals non-perturbatively, in the planar regime. Its tremendous development -- mostly driven by the breathless quest for the spectrum of scaling dimensions of single trace operators 
-- has culminated in what we can view as the alphabet for solving integrable gauge theories. 
Nevertheless, some theoretical bits are still missing, like those describing the fusion of operators in the gauge theories, that is the OPE structure constants, or the more general dual processes for which the splitting of strings holds back the integrability machinery. Of course, many important ideas have been proposed for handling theses objects and many significant results have been obtained already, on both string and gauge theory sides~\cite{oneLoopC123,roiban,MiguelKostya,tailoring1,tailoring3,Ivan2,ivan,tailoring4,Jiang:2014mja,su3,tailoringNC,ThiagoJoao,Janik:2011bd,shotasu2,Kazama:2013qsa,Klose:2012ju,Bajnok:2014sza,Hollo:2015cda,Kazama:2014sxa,Jiang:2014cya,sbsv,Vicedo:2011vn,Bajnok:2015hla,1looptwist2,volodya,Bargheer:2013faa}, but the essence of the structure constants in the integrability setup has remained elusive. 

In this letter, we shall present 
an integrability-based program for computing (non-extremal) structure constants of single-trace operators at large $N$ in the $\mathcal{N}=4$ SYM theory. Our method will appear similar, in form and spirit, to the one recently developed for scattering amplitudes in~\cite{Basso:2013vsa}. {It will feature new elementary vertices which geometrically look like hexagonal patches for the three-point functions}.
We will see precisely how these hexagons determine the three-point functions and how to set up bootstrap equations for them, akin to those developed for form factors of operators in two-dimensional integrable field theories~\cite{Watson, Cardy:2007mb}. Finally, we will present an explicit solution to these equations, valid at any coupling $g^2 = \lambda/(4\pi)^2$, and compare its predictions for the three-point functions against perturbative data at both weak and strong coupling.

\section{Hexagon program and main proposals} 

As usual, we depict the three-string interaction/three-point correlator as a pair of pants. 
Formally, a pair of pants consists of two hexagonal fundamental polygons stitched together at every other side, see figure \ref{hexagon}. 
This, in fact, is how conventional pants are sewed together along their three seams. 
\begin{figure}[t]
\begin{center}
\includegraphics[scale=0.31]{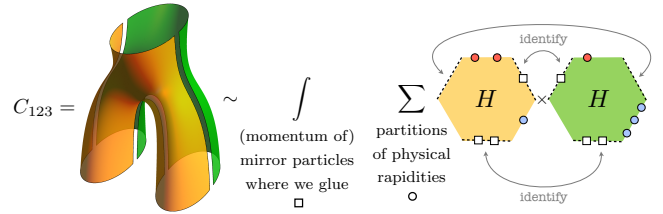}
\end{center}
\vspace{-4cm}
\caption{\normalsize  A pair of pants cut into two hexagons. Each closed spin chain operator is split into two open chains. Its excitations can end up on either half. We should sum over those possibilities.
{Stitching the hexagons back into the pair of pants amounts to integrating over all possible states at the gluing segments.}
The whole construction is reminiscent of the standard folklore: $\text{closed string}=(\text{open string})^2$.
} \la{hexagon} \vspace{-.2cm} 
\end{figure}
In the pair of pants language, each gauge invariant operator corresponds to one of the three circular cuffs and is described by a spin chain or closed string. 
%
%

When we cut the pair of pants into two hexagons we also cut each closed string into two open strings. Each of these open strings carries some of the excitations of the closed string. 
Put differently, each of two hexagons is dressed by a subset of the full set of excitations. The contribution to the three-point function for such configuration is then a product of two hexagon form factors as graphically depicted in figure \ref{hexagon}. 
An excitation can end up on either half after cutting and thus we should sum over all such possibilities, see also figure \ref{Example2}. 


When we cut the pair of pants into two hexagons we also create three new segments (the pants' seams) represented by the dashed lines in figure \ref{Example1}. When stitching the hexagons back together we should sum over all possible states living on those dashed lines. This involves integrating over the rapidities of (any number of) mirror excitations. 

Altogether we are thus lead to the proposal depicted in figure \ref{hexagon}. 
\begin{figure}[t]
\begin{center}
\includegraphics[scale=0.31]{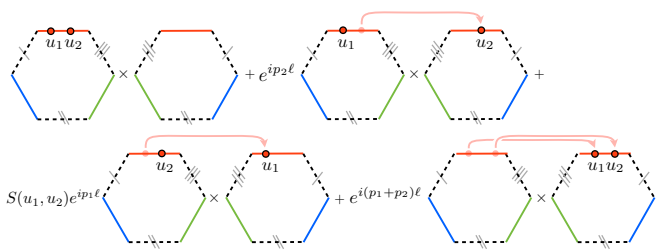}
\end{center}
\vspace{-3.6cm}
\caption{\normalsize When cutting the pair of pants into two hexagons, we should keep record of the structure of the Bethe wave function {on each cuff}. More technically, it means that, {for each chain/string}, we should sum over all possible bipartite partitions $\alpha, \bar \alpha$ of
the set of {magnons'} rapidities $\{u\}=\alpha \cup \bar \alpha$.
The weight $w(\alpha,\bar \alpha)$ dressing each term in this sum ought to take into account that for an excitation to end on the second hexagon it needs to propagate through the partial length $\ell$ of the (corresponding edge of the) first hexagon and, potentially, scatter with other excitations along the way, that is
$
w(\alpha,\bar \alpha) = \prod_{ u_j \in \bar \alpha} (e^{i p(u_j) \ell} \prod_{u_i \in \alpha \text{ with }  i > j}  S(u_j,u_i) ) 
$. 
At the level of the asymptotic wave function, this is the same cutting procedure as discussed in section 3.1 of \cite{tailoring1} at tree level. 
} \la{Example2} \vspace{-.2cm} 
\end{figure}
More concisely, we are identifying $C_{123}$ with a sort of finite volume two-point function of hexagon operators, each of them associated to the centre of mass of an hexagon. From this point of view the sum over the mirror states is the usual sum arising in form factor studies from inserting the resolution of the identity between the two operators. 
(This also provides us with a more rigorous motivation for the sum over partitions, see appendix \ref{appA}.) This point of view is backed up by a strong coupling world-sheet analysis, as discussed in appendix~\ref{ap-a}. 

This picture comes along with a notion of \textit{asymptotic} three-point function, which can be viewed as the natural counterpart of the Beisert-Staudacher asymptotic Bethe ansatz for the spectrum~\cite{Beisert:2005fw}. It applies when all lengths involved are large. 
%
More precisely, there are two kind of distances in the game: we have the length $L_i$ of each external operator and the length $l_{ij}=(L_i+L_j-L_k)/2$ ($i,j,k$ all different) of the \textit{bridge} between two given operators, see figure \ref{Example1}. The latter distances -- which are an important novelty of the three-point analysis as compared to the spectrum -- control the separation between the two hexagon operators. The asymptotic description applies when these distances are large since then we can project the sum over virtual states to the mirror vacuum. 
This is nothing but the usual clustering of correlation functions in a gapped theory. In sum, for large bridges (and thus automatically large operators) we can drop the integration over mirror particles and keep the sum over partitions of physical roots only.





At this point, we have reduced the problem of the computation of three-point functions to that of form factors of hexagon operators. The main advantage is that the former can be tackled with a (somehow) conventional integrable bootstrap, reminiscent of the one employed when studying correlators in integrable models \cite{Watson}, or branch-point twist fields \cite{Cardy:2007mb}. A similar approach was recently put forward in the context of scattering amplitudes and Wilson loops~\cite{Basso:2013vsa}. 

\begin{figure}[t]
\begin{center}
\includegraphics[scale=0.15]{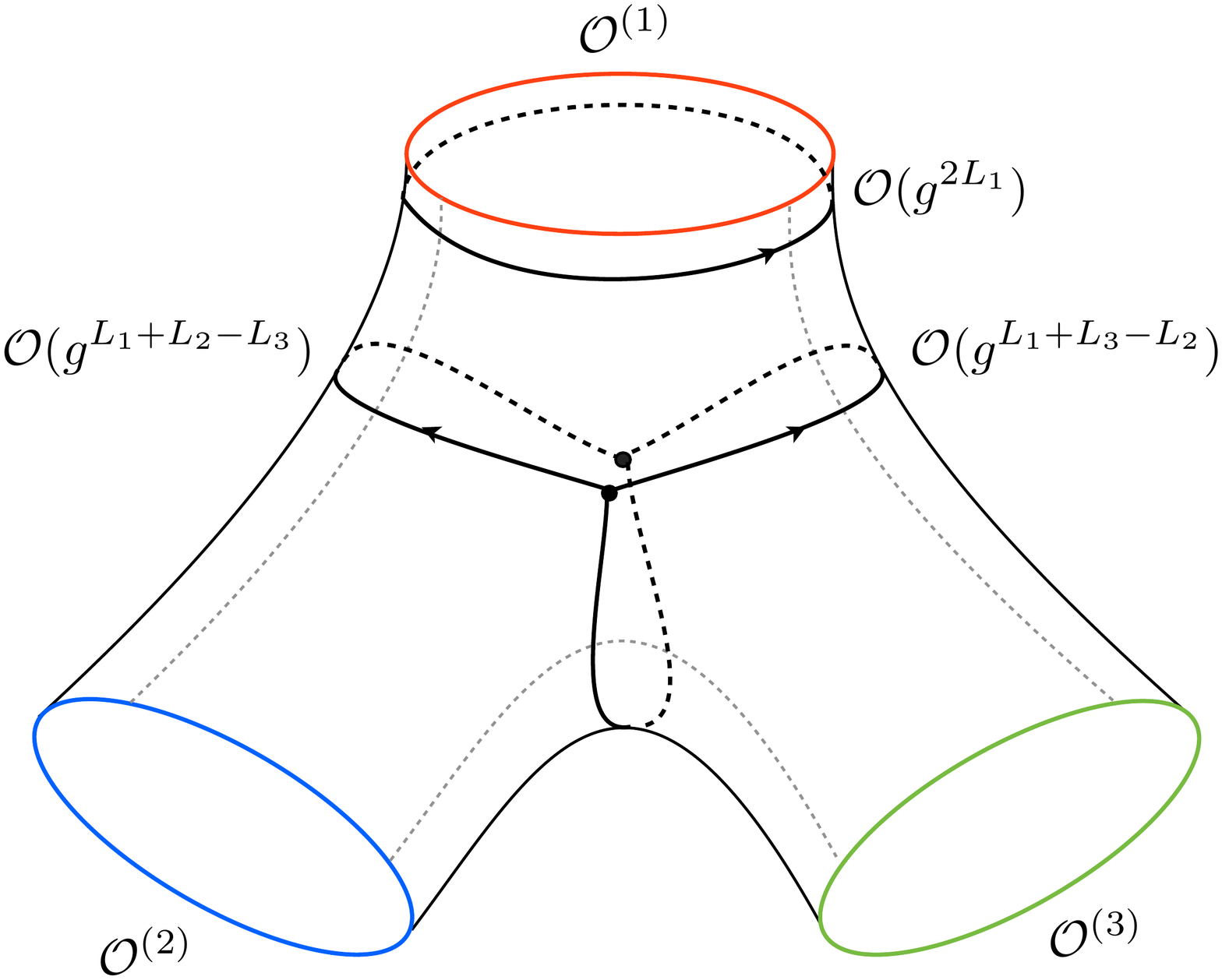}
\includegraphics[scale=0.15]{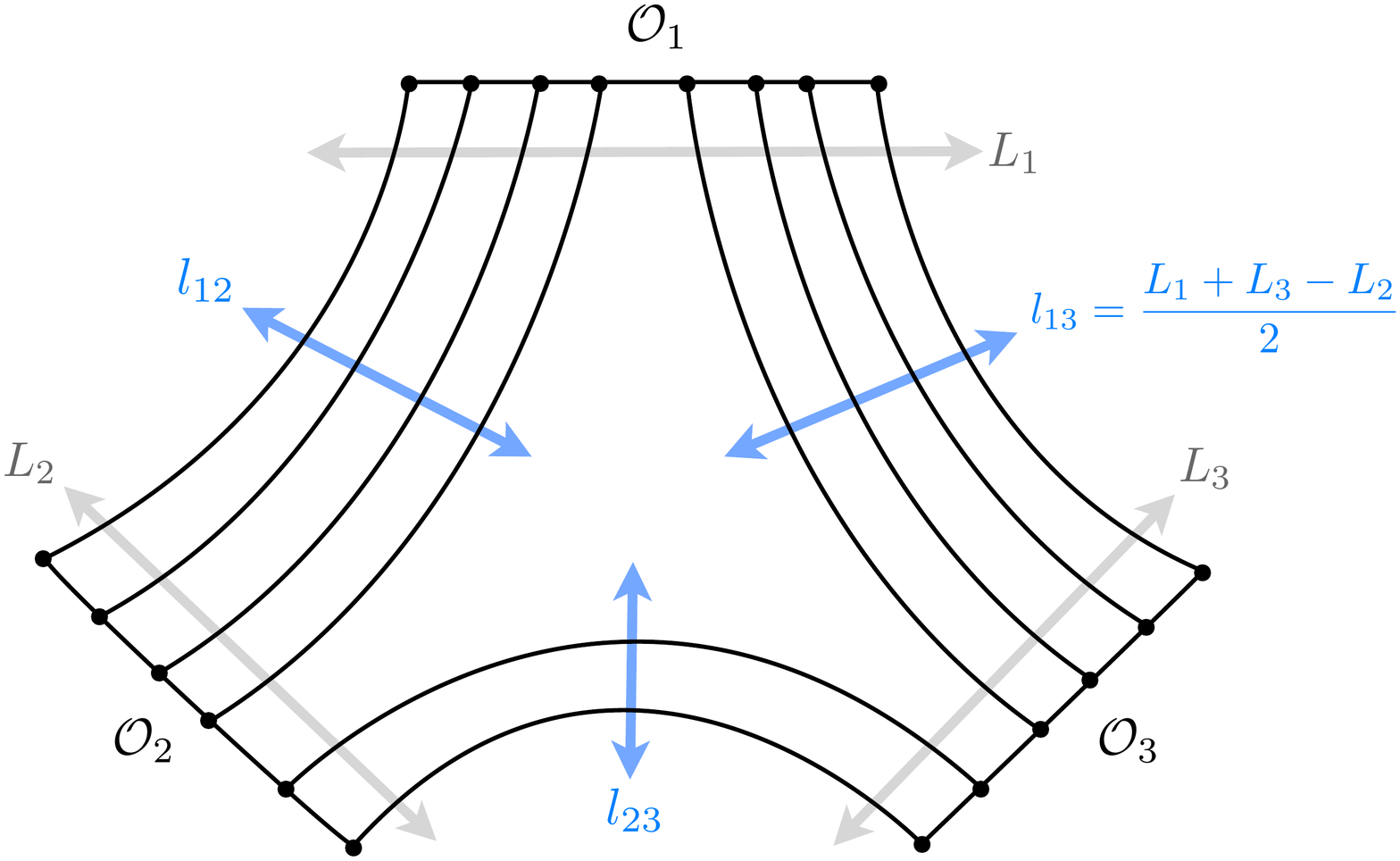}
\end{center}
\vspace{-.5cm}
\caption{\normalsize There are two kinds of finite size corrections in a three-point correlator. One is the standard wrapping corrections which encircle each of the three operators \cite{Ambjorn:2005wa}. One such correction is indicated at the top on the left. They yield corrections of order $\mathcal{O}(g^{2L_i})$ which can be dropped for long operators. Then we have a new kind of wrapping effect corresponding to virtual excitations propagating from an hexagon to another or, equivalently, from the inside to the outside of the right diagram. These effects are of order $\mathcal{O}(g^{2l_{ij}})$ and can be dropped provided the bridge length $l_{ij}$ between operators $\mathcal{O}_i$ and $\mathcal{O}_j$ is large.  In particular, it is important to note that these new wrapping corrections typically show up earlier than the conventional ones. A more detailed analysis will show that they can appear as early as 2 loops, compared to 4 loops for usual wrapping effects.  } \la{Example1} \vspace{-.2cm} 
\end{figure}

The most generic hexagon form factor is an intimidating function of six sets of variables. It depends on the physical rapidities of the three operators, as well as on the mirror rapidities of the virtual particles for each of the dashed lines in figure \ref{hexagon}. One should add on top of that the six sets of polarizations. We can, however, simplify our life by making use of so-called mirror transformations which map excitations on one edge of the hexagon to a neighbouring one, see figure \ref{ExampleMoving}. By sequencially making use of such transformations we can relate any generic hexagon to a much 
%
simpler one where all excitations live in a single physical edge. 
We denote such a creation amplitude as 
\beq
{\mathfrak h}^{A_1\dot A_1,\dots,A_N \dot A_N}(u_1,\dots,u_N)\, ,\la{ampC}
\eeq
where $A_i,\dot A_i$ are $SU(2|2)^2$ bifundamental indices parametrizing the polarization of the i-th excitation. 
%
%
%
%

Symmetry-wise the hexagon operator must be singlet under
{the $PSU(2|2)$ group of symmetries preserved by the} three BMN vacua around which we are adding excitations, see appendix \ref{symmetryAp}.
{The problem is thus constrained by the very same amount of (super-)symmetries} as was Beisert S-matrix~\cite{Beisert} or the boundary S-matrix of~\cite{DrukkerCorrea} and, {as an immediate consequence,} its solution is unique, up to an {overall} scalar factor, {for up to two magnons}.
%
%
%
%
%
%


Furthermore, combining {symmetry arguments with elementary} bootstrap considerations
hints at a simple {and natural generalization to multi-particle states. The conjecture is that the $N$-magnon hexagon amplitude~(\ref{ampC}) is \textit{exactly} given by}
\beq\label{HexA}
 {\mathfrak h}^{A_1\dot A_1\cdots}\!=\! (-1)^{\frak{f}}\prod_{i<j} {h}_{ij}   \< \ky^{\dot A_N}_{N}\!\!\dots \ky^{\dot A_1}_{1} |\,\S\,|  \ky^{A_1}_{1}\!\!\dots \ky^{A_N}_{N} \>\, ,
\eeq
where $\ky^{A} = \phi^{a}|\psi^{\alpha}$ is a state in the fundamental $SU(2|2)$ multiplet and $\S$ is Beisert $SU(2|2)$ S-matrix~\cite{Beisert} with dressing phase set to one.
($\frak{f}$ is a simple integer which accommodates for the grading~\cite{stat}.)
The multi-particle formula~(\ref{HexA}) identifies the hexagon form factor with the (factorized) scattering matrix elements up to the scalar factor $h_{ij}= h(u_{i}, u_{j})$, which is a function of two magnon rapidities. 
The latter can be constrained by crossing symmetry and argued to be given by
\beq\label{guess}
h_{12} = \frac{x_{1}^{-}-x_{2}^{-}}{x_{1}^{-}-x_{2}^{+}} \frac{1-1/x^{-}_{1}x^{+}_{2}}{1-1/x_{1}^{+}x_{2}^{+}} \frac{1}{\sigma_{12}}\, ,
\eeq
where $x^{\pm} = x(u\pm \frac{i}{2})$ are familiar Zhukowsky variables (with $u/g = x+1/x$) and $\sigma_{12}$ is (half) the BES dressing phase~\cite{Beisert:2006ez}. Accordingly, the hexagon form factor is as depicted in figure \ref{MatrixPart}, and its evaluation is straightforward, as exemplified in appendices~\ref{Offshell} and~\ref{Further}. 
\begin{figure}[t]
\begin{center}
\includegraphics[scale=0.30]{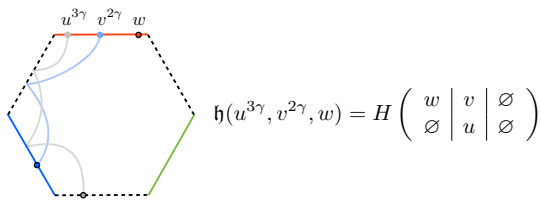}
\end{center}
\vspace{-4cm}
\caption{\normalsize A mirror transformation $\gamma: u\to u^\gamma$ moves an excitation to a neighbouring edge. As illustrated here on a simple example, we can iterate it to relate a creation amplitude $h$ with all particles at the top to the most general hexagon process $H$ where excitations can inhabit any of the six edges. 
} \la{ExampleMoving} \vspace{-.2cm} 
\end{figure}
It shows, {in the end}, some similarities with the pentagon transitions for null Wilson loops, {in that it} factorizes into a dynamical part (the product of $h$'s) and a matrix part (the S-matrix element). An important difference is that the relevant symmetry group for the null Wilson loops was just $SO(6)$ whereas here it involves
{a more sophisticated supergroup, leading, as a byproduct, to a coupling dependent matrix part.}

Relations~(\ref{HexA}) and~(\ref{guess}) finalize our proposal, which provides, \textit{in principle}, a complete non-perturbative recipe for computing structure constants of any planar gauge invariant operators in this theory. Of course, 
it is crucial to sharpen it and verify its predictions on the simplest possible examples. This is what the rest of paper is about.


\section{Properties of the Hexagon Ansatz}

In this section we elaborate on the properties of the hexagon ansatz~(\ref{HexA}).

An equivalent way of thinking about our problem is by introducing a vertex $\langle\mathfrak{h}|$ which can be contracted against three spin-chain states, like in~\cite{Kazama:2014sxa,sbsv,Jiang:2014cya}, e.g.
\beq\label{h-vertex}
\mathfrak{h}^{A\dot{A}} = \langle\mathfrak{h}|\big(|\chi^{A\dot{A}}\rangle_{1}\otimes |0\rangle_{2}\otimes |0\rangle_{3}\big)\, ,
\eeq
for a single magnon on top of the first spin chain.  We use here an invariant notation where each operator-ket is thought of as being made out of excitations on top of the \textit{same} BMN $Z$-vacuum. Implicit in there is the need to actually rotate (and translate) the kets in order to get a non-zero result compatible with $R$-charge conservation. There are several realization of these rotations, one of which is discussed in appendix~\ref{symmetryAp} and applied (up to a small twist) in the next section.

The symmetry group of each ket in~(\ref{h-vertex}) is the usual one for excitations on top of the BMN vacuum, that is the extended $PSU(2|2)^2$ introduced by Beisert in~\cite{Beisert}.
The intersection of the three symmetry groups for the three rotated vacua is a single $PSU(2|2)$, which can be thought of as a diagonal subgroup of symmetries of the BMN vacuum, as explained in appendix \ref{symmetryAp}. This group is nothing but the supersymmetrization of the obvious bosonic group $O(3)_{\textrm{Lorentz}}\times O(3)_{R-\textrm{charge}}$ that preserves 3 points in space time and 3 (generic) null vectors in `$R$-space'.  

\begin{figure}[t]
\begin{center}
\includegraphics[scale=0.31]{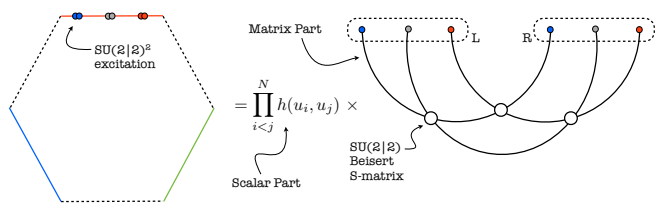}
\end{center}
\vspace{-4.2cm}
\caption{\normalsize 
The multi-particle conjecture relates the hexagon creation amplitude to a multi-particle scattering process as depicted here. 
} \la{MatrixPart} \vspace{-.2cm} 
\end{figure}

As mentioned earlier, for low number of magnons, this symmetry leaves very little freedom. For a single magnon, as explained in appendix~\ref{sym-ff}, it fixes the form factor to be~\cite{norm}
\beq\label{1-point}
{\mathfrak h}^{a\dot{a}} = \< {\mathfrak h} |\Phi^{a\dot{a}}\> = \epsilon^{a\dot{a}}\, , \qquad {\mathfrak h}^{\alpha\dot{\alpha}} = \< {\mathfrak h} |\mathcal{D}^{\alpha\dot{\alpha}}\> = \N \epsilon^{\alpha\dot{\alpha}}\, ,
\eeq
such that the only non-zero one point functions are those corresponding to so-called \textit{longitudinal} magnons, that is, the two scalars $Y = \Phi^{1\dot{2}}, \bar{Y} = \Phi^{2\dot{1}}$ and the two derivatives $D = \mathcal{D}^{1\dot{2}}, \bar{D} = \mathcal{D}^{2\dot{1}}$ polarized along the direction of the three-point function.
The relative weight $\N$ is rather arbitrary, since it absorbs the normalization freedom between states of the $PSU(2|2)^2\ltimes \mathbb{R}^3$ magnon irrep (see e.g.~(\ref{normal})). It can be fixed to $N=i$ in the commonly used string frame normalization and to $N=1$ in the spin chain one, see appendix~\ref{sym-ff}. What is important here is that the one-point function can be seen as a way of contracting the left and right part of a magnon. Put differently, it provides us with an invariant (w.r.t.~diagonal $PSU(2|2)$ subgroup) inner product
\beq\label{inn}
\<\chi^{\dot{A}}|\chi^{A}\> = {\mathfrak h}^{A\dot{A}}\, ,
\eeq 
which is the one implicitly entering in the multi-particle ansatz~(\ref{HexA}).

As alluded to before, the story repeats at two-particle level where the form factor is fixed to be of the form~(\ref{HexA}) up to the overall factor $h_{12}$ (see appendix~\ref{sym-ff} for details). As an illustration of how this formula works, we can specialize~(\ref{HexA}) to scalar indices and derive the two-scalar form factor
\beq
\begin{aligned}
{\mathfrak h}^{a\dot{a}, b\dot{b}}/h_{12} &= \<\phi^{\dot{b}}_{2}\phi^{\dot{a}}_{1}|\mathcal{S}_{12}|\phi^{a}_{1}\phi^{b}_{2}\>
\\& = A_{12}\<\phi^{\dot{b}}_{2}|\phi^{\{a}_{2}\>\<\phi^{\dot{a}}_{1}|\phi^{b\}}_{1}\> +B_{12} \<\phi^{\dot{b}}_{2}|\phi^{[a}_{2}\>\<\phi^{\dot{a}}_{1}|\phi^{b]}_{1}\> \,,
\\& = A_{12} \epsilon^{ a \dot b}\epsilon^{b \dot a} +\tfrac{1}{2}(A_{12}-B_{12})\epsilon^{ a b}\epsilon^{\dot a \dot b} \,,
\end{aligned}
\eeq
after $\mathcal{S}$-acting as in~\cite{Beisert:2006qh} and contracting the resulting bra-ket using~(\ref{inn}). We could proceed similarly for derivatives and fermions. In particular, we would find that the form factor ratio between two $Y$'s and two $D$'s is simply given by
\beq\label{ADratio}
h_{Y_{1}Y_{2}}/h_{D_{1}D_{2}} = -A_{12}/D_{12} = \frac{x_{1}^{-}-x_{2}^{+}}{x_{1}^{+}-x_{2}^{-}}\, ,
\eeq
upon evaluation of the $A$ and $D$ component of the S-matrix in the spin-chain frame \cite{Beisert, Beisert:2006qh} (with $N=1$). A similar expression can be found in the string frame \cite{reviewHuge, AF-string}, as detailed in appendices~\ref{sym-ff} and \ref{mapping}.
\begin{figure}[t]
\begin{center}
\includegraphics[scale=0.31]{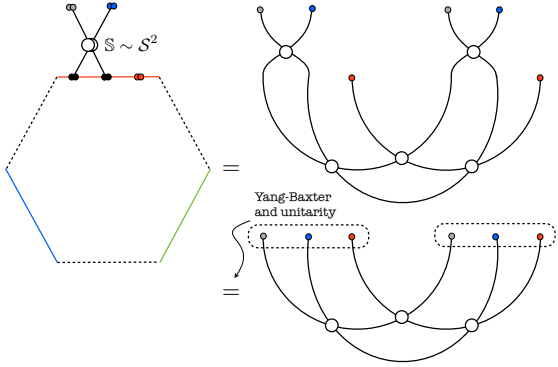}
\end{center}
\vspace{-2cm}
\caption{\normalsize Because the S-matrix is given by a product of two identical left and right S-matrices, Watson equation is automatically satisfied by the multi-particle ansatz. 
} \la{fwatson} \vspace{-.2cm} 
\end{figure}

Equation~(\ref{ADratio}) illustrates the power of the symmetry. Heuristically, the reason why this symmetry appears so efficient here is that the left and right part of the magnon behave, respectively, as particle and antiparticle w.r.t.~the diagonal subgroup (see appendices \ref{symmetryAp}, \ref{sym-ff} and \ref{cross-app} for details).
Upon crossing of the right part, the set up becomes thus identical to the one considered by Beisert, which is why its (unique) S-matrix is recovered. The only difference resides in the scalar factor of course, that needs not be unitary in our case. \\

The ansatz~(\ref{HexA}) clearly stands as the most natural generalization to many particles, but, as should be clear by now, is not fixed by diagonal $PSU(2|2)$ symmetry alone, for three or more magnons. We shall now provide evidence for it and unveil some of its nice features.

To start with, it is easy to see that the hexagon vertex~(\ref{HexA}) is automatically preserved by the action of the S-matrix. This is the content of the familiar Watson equation for form factors :
\beq\label{watson}
\<{\mathfrak h} | (\mathbb{S}_{ii+1}-\mathbb{I})|\ldots \chi^{A_{i}\dot{A}_{i}}_{i}\chi^{A_{i+1}\dot{A}_{i+1}}_{i+1}\ldots  \> = 0\, ,
\eeq
with $\mathbb{S} = S^{0}(-1)^{\dot{F}}\mathcal{S}\, \dot{\mathcal{S}}(-1)^{F}$ the $SU(2|2)^2$ S-matrix~\cite{trivial}.
As depicted in figure~\ref{fwatson}, this equation immediately follows from the structure of the ansatz and the unitarity of the S-matrix, assuming the scalar factor $h_{12}$ in~(\ref{HexA}) verifies
\beq\label{Watscal}
h_{12}/h_{21} = S_{12}^{0} = \frac{x_{1}^{+}-x_{2}^{-}}{x_{1}^{-}-x_{2}^{+}}\frac{1-1/x^{-}_{1}x_{2}^{+}}{1-1/x^{+}_{1}x^{-}_{2}}\frac{1}{\sigma^{2}_{12}}\, ,
\eeq
where $S_{12}^{0}$ is the spin chain scattering matrix in the $\mathfrak{sl}(2)$ subsector.\\

\begin{figure}[t]
\begin{center}
\includegraphics[scale=0.31]{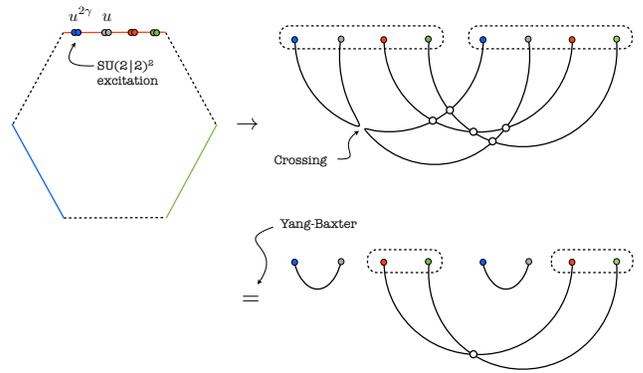}
\end{center}
\vspace{-2cm}
\caption{\normalsize A particle/anti-particle pair automatically decouple from the multi-particle ansatz. This follows from crossing and Yang-Baxter for Beisert S-matrix. 
} \la{dec} \vspace{-.2cm} 
\end{figure}

A second piece of evidence comes from the so-called decoupling condition. Indeed, form factors typically show kinematical poles whenever a particle-antiparticle pair (with zero net energy and momentum) decouples from the rest of the state. The precise condition can take different form for different type of operators, see e.g.~\cite{Watson} for local operators. In our case, it is the form appropriate to the class of non-local operators considered in~\cite{Basso:2013vsa,Cardy:2007mb,Basso:2014jfa} that is relevant. Accordingly, we shall insist that our form factors develop a pole, whenever such a pair is formed, with a residue that is simply proportional (i.e.~with no interference term) to the form factor for the subprocess in which the pair has been removed. Since our ansatz is directly written in terms of the S-matrix, this condition is easily implemented and almost immediately satisfied. Indeed, on any such a pair, the S-matrix develops a pole with residue mapping to a singlet of the symmetry algebra. This means that we can write
\beq\label{eleven}
\S \big|\chi^{A}_{1}\chi^{B}_{2} \prod_{\textrm{rest}}\chi_{j}\big>_{\textrm{pole}\, (12)} \propto \S_{\textrm{rest}}\prod_{\textrm{rest}}\S_{2j}\S_{1j}\big|\prod_{\textrm{rest}}\chi_{j} \times \mathrm{1}_{21}\>
\eeq
where the equation holds at the level of the pole in the $(12)$ channel and $\big|\mathrm{1}_{21}\big>$ is Beisert singlet~\cite{Beisert:2006qh} built out of the particle-antiparticle pair $(12)$. The decoupling is then guaranteed if the scattering between the arbitrary excitation $j$ and the pair $(12)$ is trivial, i.e.~equal to $1$, see figure~\ref{dec}. This condition is the same as Janik crossing equation for the S-matrix~\cite{Janik:2006dc} as derived by Beisert in~\cite{Beisert:2006qh}, except that the scalar factor there should now be replaced by~$h_{12}$. In other words, for our factorized ansatz, the decoupling condition boils down to the crossing equation
\beq\label{Cross}
h(u_{1}^{2\gam}, u_{2})h(u_{1}, u_{2}) = \frac{x_{1}^{-}-x_{2}^{-}}{x_{1}^{-}-x_{2}^{+}}\frac{1-1/x^{+}_{1}x_{2}^{-}}{1-1/x^{+}_{1}x_{2}^{+}}\, ,
\eeq
where $u_{1}^{2\gam}$ denotes the crossed rapidity (corresponding to the swapped Zhukowsky variables $x^{\pm}_{1} \rightarrow 1/x^{\pm}_{1}$).

The simplest solution we could find to both~(\ref{Watscal}) and~(\ref{Cross}) is the one we reported in~(\ref{guess}).~\cite{comment}\\


As mentioned earlier, starting from a form factor with all excitations on the same edge, we can reach any configuration by means of mirror or crossing transformations. These transformations are simpler to implement in the string frame, as explained in detail in appendix~\ref{cross-app}. Crossing, in this case, simply amounts to performing twice the analytical continuation $\gam$ mentioned earlier along with a suitable re-arranging of the indices of the excitations. Precisely, in the string frame,
\beq\label{crossingrule}
\big|\chi(u^{2\gamma})\big\rangle_{1}\left|0\right>_{2}\left|0\right>_{3} = \left|0\right>_{1}\big|-\tilde{\chi}(u)\big>_{2}\left|0\right>_{3} \, , 
\eeq
with
\beq\label{dotting}
\tilde{\chi}^{A\dot{B}} = \chi^{B\dot{A}} \, ,
\eeq
while, for comparison, in the spin chain frame, one might also have to multiply by extra momentum-dependent factors, as explained in detail in appendix \ref{mapping}, depending on the excitation.
In either frame, the crossing transformation for derivatives is straightforward and simply given by (\ref{crossingrule}), such that e.g. the hexagon process with one $D$ on chain $1$ and its conjugate on chain $2$ reads
\beq\label{direct-t}
h_{D|\bar{D}}(u|v) \equiv - h_{DD}(v^{2\gam}, u)\, . 
\eeq
(Explicit expressions for $h_{DD}$ with crossed rapidities are given in appendix \ref{crossingFormulae}.) For identical particles on the two edges, the transition $h_{X|X}(u|v)$ must have a pole at $u = v$ in agreement with~(\ref{eleven}), see~(\ref{mu-def}) below for details. The residue at this pole is controlled by the measure $\mu_{X}(u)$,
\beq\label{mu-def}
\underset{v=u}{\operatorname{{\rm res}}} \,\,h_{X|X}(u|v)= \frac{i}{\mu_{X}(u)}\, .
\eeq
Recall that in this limit the excitation decouples from the rest, such that the measure just originates from the overlap of the in and out states. It governs the (asymptotic) normalization of the one-particle state and is an essential ingredient for a properly normalized three-point function, as illustrated in section~\ref{all-loop-asy}. A similar measure was introduced in the pentagon approach for null Wilson loops, where it was also used for `stapling' together two pentagons overlapping on a square. The analogous gluing procedure for the hexagons is depicted in figure~\ref{Stapling} below.

%

 Let us conclude this section by a brief comment.
%
It so happens that the two hexagons in the three-point function are not absolutely identical. Instead, their form factors differ by $(-1)^M$ where $M$ is the total number of magnons.  Originally, we found this simple rule by staring at weak and strong coupling data. (It results in the $(-1)^{|\alpha|}$ signs that will pop up everywhere below.) A posteriori, we also came up with a few, plausible but non rigorous, geometrical and kinematical explanations for it.
Hopefully, an odd number of such arguments hold. It would be very nice to better understand the origin of this empirical rule, which is responsible for restoring certain symmetries of the structure constants. 

%

\section{Simple weak coupling checks} \la{simpleWeak}

The simplest way to access to the elementary hexagon transitions is by considering processes
with a single excitation $\chi_\text{top}$ with rapidity $u$ on one operator $\mathcal{O}_1$ and another single excitation $\chi_\text{bot}$ with rapidity $v$ on the operator $\mathcal{O}_2$. We shall refer to these operators as top and bottom, respectively, in accord with figure~\ref{TreeLevel}.
Of course, to probe such simple configurations we have to relax the zero momentum condition for each operator. We shall nevertheless still keep the Bethe ansatz conditions 
\beq
e^{i p L_1} = e^{i q L_2}=1  \,, \la{BAEexample}
\eeq
where $p=p(u)$ and $q=p(v)$.  The third operator $\mathcal{O}_3$ has no excitations; it is a BPS operator. 


Upon cutting the pants in two, the single excitations on the top or in the bottom can end up on either hexagon. As mentioned above, when an excitation ends on the second hexagon we are instructed to include an extra minus sign. In sum, we expect to find
\beqa
\!\!\!\!\!\!\!\!C_{123}^{\bullet\bullet\circ} &\propto& h_{\chi_\text{top}|\chi_\text{bot}}(u|v) + e^{i p \ell_{31}+i q \ell_{12}} h_{\chi_\text{bot}|\chi_\text{top}}(v|u) \nn\\
\!\!\!\!\!\!\!\!&&\!\!\!\!\!\!\!\!\!\! -e^{i q \ell_{12}} h_{\chi_{\text{top}}}(u)h_{{\chi}_{\text{bot}}}(v)-e^{ip \ell_{31}}  h_{\chi_{\text{top}}}(u)h_{{\chi}_{\text{bot}}}(v)  \,.\nn\la{expect}
\eeqa
All these quantities can be neatly predicted by starting with our finite coupling proposal for the creation amplitude and crossing one of the particles down by a sequence of mirror transformations.
For the case of a longitudinal scalar $Y$ and/or its conjugate $\bar Y$, for instance, we get
\beq
h_{Y|Y}= 1-\frac{i}{u-v} \, , \,\,\,\,\, h_{Y|\bar{Y}}= -1 \, , \,\,\,\,\, h_{Y} = - h_{\bar Y}=1 \,,\la{expectY} 
\eeq
to leading order at weak coupling. We will now compare these predictions against explicit tree level computations.

\begin{figure}[t]
\begin{center}
\includegraphics[scale=0.32]{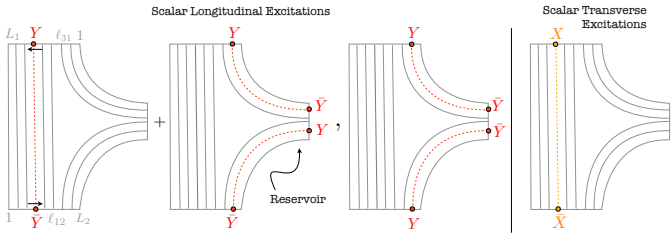}
\end{center}
\vspace{-4cm}
\caption{\normalsize
We can directly probe the hexagon transition $h(u|v)$ by inserting excitations on two of the legs of the three-point function. The excited operators (here depicted at the bottom and top) are then conveniently seen as the future and past ends of a cylinder. This cylinder is coupled to a rotated BPS vacuum (the third operator on the right) which acts as a reservoir:
Longitudinal excitations can be absorbed by the reservoir while transverse excitations cannot and thus must be connected directly from top to bottom. 
} \la{TreeLevel} \vspace{-.2cm} 
\end{figure}

With this in mind, we introduce a particular realization of the setup above to which we refer as the \textit{reservoir picture}, see figure~\ref{TreeLevel}. 
In this picture, the first operator $\mathcal{O}_1$ is taken to be of the form $\Tr(Z\dots Z \chi_{top} Z \dots Z) + \dots$ and is located at $x=0$, the second operator $\mathcal{O}_2$ reads $\Tr(\bar Z\dots \bar Z \tilde{\chi}_{bot} \bar Z \dots \bar Z) + \dots$, with $\tilde{\chi}$ as defined in~(\ref{dotting}), and is located at $x=\infty$ and, finally, the third operator is inserted at $x=(0,1,0,0)$ and reads
\beq
\mathcal{O}_3 = \Tr(Z+\bar Z + Y - \bar Y)^{L_3} \la{reservoirOp}\,.
\eeq
Note that despite the slightly unorthodox appearance, this is a simple (rotated) BPS operator on the same footing as $\Tr(Z^{L_3})$. The fact that it is rotated (it has to be otherwise one can not properly contract it with the other two in a non-extremal setup) is, however, quite important and justifies the name of the \textit{reservoir}.
We see, for instance, that some excitations of the operators $\mathcal{O}_1$ or $\mathcal{O}_2$ can be absorbed (i.e. Wick contracted with) by this reservoir. 
These are precisely the longitudinal excitations introduced earlier. The remaining excitations -- namely fermions, transverse derivatives and other scalars -- are what we call the \textit{transverse} excitations. If we insert a transverse excitation in the top (a scalar $X = \Phi^{1\dot{1}}$ say) we ought to introduce its complex conjugate (i.e.~$\bar{X} = -\Phi^{2\dot{2}}$) at the bottom or else the result will be zero. In short, the reservoir setup is a realization of the rotations alluded to below (\ref{h-vertex}) and further detailed in appendix~\ref{symmetryAp}, combined with an inversion sending the second operator to infinity.

There are clearly two fundamental quantities arising from the tree-level Wick contractions (see figure \ref{TreeLevel}) in terms of which we can construct any process for different excitations $\chi_\text{top}$ and $\chi_\text{bot}$.
Namely, we can either connect an excitation in the bottom with one at the top by a propagator, thus generating a contribution  
\beq
C^\text{direct}=\sum_{n=1}^{\ell_{12}} e^{i p (L_1-n+1) + i q n} \, , \la{s1}
\eeq
or, if the excitation is longitudinal, it can connect directly to the reservoir. When a top/bottom excitation with momentum $p$/$q$ is connected with the reservoir it generates a corresponding contribution of the form 
\beq
C^\text{reservoir}_\text{top}= \sum_{n=1}^{\ell_{31}} e^{i p n} \, , \,\,\,\,\,\,\, C^\text{reservoir}_\text{bottom}= \sum_{n=\ell_{12}+1}^{L_2} e^{i q n} \,. \la{s2}
\eeq
It is now a simple matter of appropriately adding up these contributions. 

As a warmup, consider a single excitation $Y$ at the top. In this case the structure constant comes from a single term
\beq\label{twenty3}
C^\text{reservoir}_\text{top}  =  \mathcal{N}(p)(1- e^{i p \ell_{31}})\, ,
\eeq
where $\mathcal{N}(p) = 1/(e^{-ip}-1)$ reflects the specific plane wave normalization we are using here. Since this factor plays no role in the following discussion, it will always be implicitly factored out and disregarded. More interesting is the relative minus sign between the two terms in the numerator in~(\ref{twenty3}), which arises trivially from the geometric nature of the series. This is the simplest instance of the minus signs alluded to in the end of the previous section. Apart from this sign and the normalization factor, the result~(\ref{twenty3}) is just the sum of two one point hexagon functions, as expected.

Consider now the more interesting case of a direct transition $Y\to Y$ corresponding to $\chi_\text{top}=\chi_\text{bot}=Y$. As explained above, in the reservoir picture, this configuration corresponds to inserting a scalar $Y$ at the top and its conjugate $\bar Y$ at the bottom (on top of the $Z$ and $\bar{Z}$ vacuum, respectively). These excitations can either be directly connected by a propagator or they can be separately absorbed by the reservoir, see first two pictures in figure \ref{TreeLevel}. The latter contribution ought to pick an additional minus sign w.r.t.~the former since in the reservoir operator  (\ref{reservoirOp}) there is an explicit relative minus sign between the $Y$ and the $\bar Y$. Altogether we find therefore 
\beq
\left. C_{123}^{\bullet\bullet\circ} \right|_{Y \to Y}\propto C^\text{direct} - C^\text{reservoir}_\text{top} \times C^\text{reservoir}_\text{bottom} \,.
\eeq
Evaluating the simple geometric sums in (\ref{s1}) and (\ref{s2}) and simplifying the outcome using Bethe equations (\ref{BAEexample}) we finally get
\beqa
\left. C_{123}^{\bullet\bullet\circ} \right|_{Y \to Y} &\propto& (1-\frac{i}{u-v}) + e^{i p \ell_{31}+i q \ell_{12}} \times (1-\frac{i}{v-u})   \nn\\
&&\!\!\! - e^{i q \ell_{12}} \times 1-e^{ip \ell_{31}} \times  1  \,,\la{first}
\eeqa
in perfect agreement with our expectations (\ref{expectY}).
Next, we could look at the transition $Y\to \bar{Y}$ which amounts to inserting a field $Y$ at both the top and bottom. At tree level we cannot contract these fields by a propagator, of course, but they can still be independently absorbed by the reservoir.
Accordingly, we must have
\beq
\left. C_{123}^{\bullet\bullet\circ} \right|_{Y \to \bar{Y}}\propto  C^\text{reservoir}_\text{top} \times C^\text{reservoir}_\text{bottom} \, ,
\eeq
or equivalently
\beqa
\left. C_{123}^{\bullet\bullet\circ} \right|_{Y \to \bar{Y}}&\propto& (-1) + e^{i p \ell_{31}+i q \ell_{12}} \times (-1)   \nn\\
&&\!\!\! -\, e^{i q \ell_{12}} \times (-1)-e^{ip \ell_{31}} \times  (-1) \, ,\la{second}
\eeqa
which is again in perfect agreement with our expectations~(\ref{expectY}). Consider finally the case where the scalar is a transverse excitation $X$. Then only the direct propagator contributes and
\beq
\left. C_{123}^{\bullet\bullet\circ} \right|_{X \to \bar{X}}\propto  C^\text{direct} \,.
\eeq
We can therefore simply add up the right hand sides of (\ref{first}) and (\ref{second}) to get 
\beqa
\left. C_{123}^{\bullet\bullet\circ} \right|_{X \to \bar{X}}&\propto& -\frac{i}{u-v} + e^{i p \ell_{31}+i q \ell_{12}} \times \frac{-i}{v-u}  + \nn\\
&&\!\!\! e^{i q \ell_{12}} \times 0+e^{ip \ell_{31}} \times  0  \,.\la{third}
\eeqa
From the zeros in the second line, we learn that there is no one point function for such transverse excitations, in line with the symmetry analysis, while from the first line we derive that
\beq
h_{X|\bar{X}} =  -\frac{i}{u-v}\, ,
\eeq
which perfectly matches with what we would extract out of our ansatz.

It is not so much complicated to extend the previous analysis to loop level. It suffices to decorate the tree level Wick contraction by the insertion of the one loop scalar dilatation operator at the splitting points \cite{oneLoopC123}. Using the single excitation case to normalize our three-point function, we can then promptly read off the first quantum correction to various $h$-transitions. In this way we would find
\beqa
&&h_{Y|Y}(u,v)=\frac{u-v-i}{u-v}  \Big[1+\frac{g^2}{(u^2+\tfrac{1}{4})(v^2+\tfrac{1}{4})} + O(g^4) \Big] \,,\nn \\
&&h_{Y|\bar Y}(u,v)=- 1 +O(g^4)\,, \nn \\
&&h_{X|\bar X}(u,v)=h_{Y|Y}(u,v)+h_{Y|\bar Y}(u,v) \, ,\la{expansions}
\eeqa
in perfect agreement with the one loop expansion of our all loop conjectures. Note in particular that the function $h_{Y|Y}(u,v)$ is nothing but the loop corrected $f(v,u)$ in~\cite{tailoring4}. 

We further extended the tree level analysis above to all other magnon excitations always finding a perfect match with our predictions. (See also appendices \ref{Offshell}, \ref{Further} and \ref{Impossible} for more involved checks at weak coupling.) For fermions, for instance, we could use the recent results for the splitting insertions of \cite{ThiagoJoao} to test their transition up to one loop. In the end, it is quite amusing that so much of the full result can already be checked with such simple computations involving no more than one particle in the top and another in the bottom!

\section{$SU(2)$ and $SL(2)$ Asymptotic All Loop Conjecture}\label{all-loop-asy}

Putting together all the ingredients in our main proposal we will now write down
the final prediction for the asymptotic all-loop structure constants for some simple (so-called rank one) sectors involving one non-trivial operator and two protected operators. Our proposal will extend the one loop results of \cite{tailoring1,tailoring4} and \cite{tailoringNC} -- for $SU(2)$ and $SL(2)$ correlators respectively -- to all loops. Both can be described at once, by replacing $h$ below by $h_{DD}$ for the $SL(2)$ case or by $h_{YY}$ for the $SU(2)$ setup, {and similarly for all other dynamical quantities}. To avoid unimportant combinatorial normalization factors we present here the ratio between the structure constant and a structure constant of three BPS operators with the same lengths. Our conjecture then concisely reads
\beq
\(\frac{C_{123}^{\bullet \circ \circ}}{C_{123}^{\circ \circ \circ}}\)^2 = \frac{\prod_{k=1}^S \mu(u_k)}{\det\, \partial_{u_i} \phi_j \prod_{i<j}S(u_i, u_j)} \times \mathcal{A}^2\, , \la{rank1}
\eeq
where $S$ is the number of magnons in the non-protected operator (scalars $Y$ in the $SU(2)$ case or derivatives $D$ for the $SL(2)$ setup). It makes use of the measure $\mu(u)$ defined through the residue
of the direct transition as in~(\ref{mu-def}), of the Gaudin norm (here in rapidity $u$ and not momentum $p$ space) defined with
\beq
e^{i \phi_j} \equiv  e^{ip_{j}L_{1}} \prod_{k\neq j} S(u_{j}, u_k)\, , \la{BAE}
\eeq
and of a product of scattering phases. So combined, these quantities permit the conversion between infinite- and finite-volume normalizations, see e.g.~\cite{Pozsgay:2009pv,Bajnok:2015hla} and references therein for relativistic systems.

Finally, and most interestingly, we have the sum over the hexagon processes, which can be cast into the form
\beq
\mathcal{A} = \prod_{i<j}h(u_{i}, u_{j})\sum_{\alpha\cup \bar{\alpha}={\bf u}} (-1)^{|\bar{\alpha}|} \prod_{j\in \bar{\alpha}}  e^{ip_{j}\ell} \prod_{i\in \alpha, j\in \bar{\alpha}} f(u_{i}, u_{j})  \la{calAformula}
\eeq
where we sum over all bipartite partitions $\alpha\cup\bar\alpha =\{u_i\}$ of the set of Bethe rapidities, with $f(u, v) = 1/h(u, v)$ and with the splitting length $\ell = \ell_{31}$. A few more comments on this expression and on the connection to previous expressions in the literature for such correlators at low loop order are presented in appendix~\ref{complementCalA}. 

%

\section{{Leading finite-size corrections}}\label{FSsection}

The asymptotic formula~(\ref{rank1}) for $C_{123}^{\bullet \circ \circ}$ can be systematically improved by incorporating finite-size corrections.
In our picture, these ones come about when particles get exchanged between the two hexagons.  At weak coupling, the heavier the exchanged state is the later it contributes in perturbation theory. The leading finite-size corrections are thus obtained when a single particle is passing through one of the three mirror channels along which we glue the two hexagons together, see figure~\ref{Stapling}. This amounts to replacing $\mathcal{A}$ in (\ref{rank1}) by 
\beq
\mathcal{A} \to \mathcal{A} + \delta\mathcal{A}_{{12}} + \delta\mathcal{A}_{{23}}+ \delta\mathcal{A}_{{31}} \,,  
\eeq
where the corrections $ \delta\mathcal{A}_{{12}}$ and $ \delta\mathcal{A}_{{31}}$ are associated to the neighboring bridges while $\delta\mathcal{A}_{{23}}$ is the contribution from the opposing bridge. As we will see they yield very different contributions. Nevertheless, regardless of which one of these channels we consider, the one-particle contribution always takes the form
\beq\label{calAwrapping}
\delta\mathcal{A} =\sum_{a\geqslant 1}\int \frac{du}{2\pi}  \mu^{\gamma}_{a}(u)\times \(\frac{1}{x^{[+a]}x^{[-a]}}\)^{\ell}\times \textrm{int}_{a}(u|\{u_{i}\})\, ,
\eeq
with $\ell$ the size of the corresponding bridge. The recipe is that we ought to sum over the tower of bound states ($a=1, 2, ...$) and integrate over the rapidity $u$ of the mirror excitation with the help of the mirror measure $\mu^{\gamma}_{a}(u) \equiv \mu_{a}(u^{\gamma})$. For a bound state of $a$ derivatives, this measure reads explicitly as~\cite{fusion}
\beq\label{bs-measure}
\mu_{a}(u^{\gamma}) = \frac{a (x^{[+a]}x^{[-a]})^2}{g^2(x^{[+a]}x^{[-a]}-1)^2((x^{[+a]})^2-1)((x^{[-a]})^2-1)}\, ,
\eeq
with $x^{[\pm a]} = x(u\pm i\frac{a}{2})$, and it differs by an overall sign from the mirror measure for the bottom (or $Y$-like) component in the same supermultiplet. Thanks to the diagonal $O(3)\times O(3)$ symmetry, these are the only cases we have to consider; that is, for the type of three-point functions under study, the mirror excitation in~(\ref{calAwrapping}) can only be one of the longitudinal bosons $D, \bar{D}, Y, \bar{Y}$ or one of their homologues in the bound state multiplet. For sake of clarity, the corresponding flavour summation is implicitly understood in~(\ref{calAwrapping}) and subsumed into the integrand $\textrm{int}_{a}$ as explained below.

\begin{figure}[t]
\begin{center}
\includegraphics[scale=0.35]{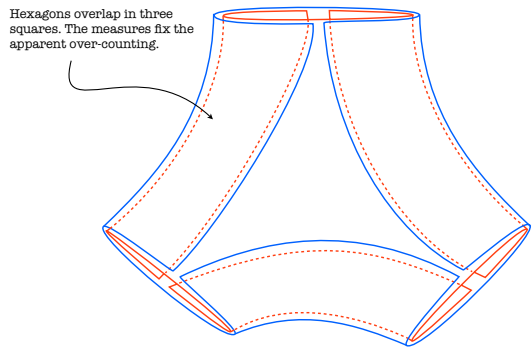}
\end{center}
\vspace{-2.8cm}
\caption{\normalsize 
We can glue two hexagons by overlapping them along three squares. The integration measures $\mu$ take into account the propagation of states in these squares and remove the naive over-counting. This picture is akin to the way pentagons are glued together into scattering amplitudes \cite{Basso:2013vsa}.
} \la{Stapling} \vspace{-.2cm} 
\end{figure}

The second term in the integrand in (\ref{calAwrapping}) is the usual damping factor $e^{-\ell E_{a}(u)}$ which accommodates for the propagation through a distance $\ell$ of a mirror particle with energy $E_{a} = \log{x^{[+a]}x^{[-a]}}$. 

Finally, the last remaining ingredient is the integrand $\textrm{int}_{a}(u|\{u_{i}\})$ which collects the hexagon form factors for creation and annihilation of the mirror excitation in the background of real magnons $\{u_i\}$. 


Based on this data alone, we can already estimate the finite size correction at weak coupling. Since in this limit $x^{[\pm a]}\sim (u\pm ia/2)/g$, the integral~(\ref{calAwrapping}) is found to be parametrically of order $O(g^{2+2\ell})$, assuming an integrand $\textrm{int}_{a}(u|\{u_{i}\})$ of order $O(1)$. The virtual effects captured by~(\ref{calAwrapping}) thus start generically at $(\ell+1)$ loops for a bridge of size $\ell$, which means, in particular,  two loops for the smallest (non-extremal) bridge, with $\ell = 1$.

The integral~(\ref{calAwrapping}) mimics in many ways the L\"usher formula for the spectrum~\cite{Bajnok:2008qj}. In fact, the parallelism goes even further since its integrand also features a transfer matrix. To see that, we need to look more closely at what is running inside the integrand. For the sake of clarity, we shall consider the $a=1$ contribution to the adjacent mirror channel (obtained by a single $\gamma$ move away from the excited cuff of the pants). In this case, we get 
\beqa
&&\textrm{int}^{\gamma}(u|\{u_{i}\}) = \sum_{\alpha \cup \bar \alpha = \{u_i\}} w(\alpha,\bar \alpha) (-1)^{|\alpha|} \times \la{integrand} \\ &&\qquad \sum_{X=D, \bar{D}, Y, \bar{Y}} (-1)^{f_X}h_{XD\dots D}(u^{\gamma}, \alpha})h_{D\dots D\bar{X}} ({\bar{\alpha}, u^{-\gamma}) \nn
\eeqa
where $w(\alpha,\bar \alpha)$ is the splitting factor given in figure \ref{Example2}.
Of course, this is a minor modification of the asymptotic result (\ref{calAformula}); indeed, if we replace $X$ in the second line by the vacuum, the two hexagons become fully decoupled and we recover precisely the asymptotic result, see also appendix \ref{complementCalA}. 

\begin{figure}[t]
\begin{center}
\includegraphics[scale=0.31]{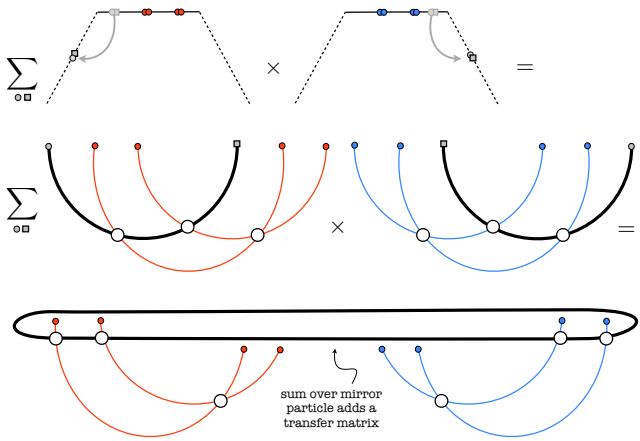}
\end{center}
\vspace{-0.9cm}
\caption{\normalsize When sticking the hexagons back into the pair of pants we should sum over all possible mirror excitations. For a single particle, the corresponding sum over particle flavour neatly reconstructs a transfer matrix multiplying the asymptotic result. 
} \la{transfer} \vspace{-.2cm} 
\end{figure}

The sum over fundamental particles $X$ in the second line of (\ref{integrand}) is simply what we expect for the production of a pair $X(u^{\gamma})\bar{X}(u^{-\gamma})$ of mirror particles, which are respectively absorbed by the $\alpha$- and $\bar{\alpha}$-decorated hexagon. This is the leading effect coupling the two hexagons. 
The grading by $f_{X} = 1, 0$ for $D$'s and $Y$'s, respectively, originates from the relative sign between their corresponding measures. This structure is clearly reminiscent of the one for a transfer matrix. 
 In fact, up to an overall dynamical factor (which depends on the partition), the $X$ (or flavours) sum precisely reconstructs the $\mathfrak{su}(2|2)$ transfer matrix $T(u^{-\gamma}|\{u_i\})$ obtained by tracing the fundamental $\textbf{2}|\textbf{2}$ irrep around the Bethe state, see figure~\ref{transfer}. Since the latter is partition independent, it can be factored out in~(\ref{integrand}) which can then be schematically written as
 \beq
\textrm{int}^{\gamma}(u|\{u_{i}\}) \propto T(u^{-\gamma}|\{u_i\})\, .
\eeq
What is important here is that it is well-known that this transfer matrix is actually suppressed at weak coupling, thanks to supersymmetry; i.e.~$T(u^{-\gamma})\sim g^2$. Since the overall dynamical factor is, on the other hand, of order $O(1)$, we conclude that finite size effects are delayed up to $(\ell+2)$ loops in the adjacent channels (i.e.~for $\ell=\ell_{13}$ or $\ell=\ell_{12}$). Roughly speaking, we could say that finite size effects in these channels behave as half-wrapping effects, since the latter ones feature a product of left and right transfer matrices and are suppressed by two powers of~$g^2$.

This supersymmetric delay is not realized however in the mirror channel which is facing the excited state. This one is obtained by considering mirror excitation standing $3\gamma$ away from it. In this case, the end result for the integrand takes a very simple form. Generalizing to bound states and explicitly performing the sum over partitions, we can write it concisely as
\beq
\textrm{int}^{3\gamma}_{a}(u|\{u_{i}\}) = \mathcal{A}\times\frac{(-1)^a T_{a}(u^{\gamma})}{\prod_{i}h_{D_{a}D}(u^{\gamma}, u_{i})}\, , \la{voila1}
\eeq
where $\mathcal{A}$ is the asymptotic piece written before, see~(\ref{calAformula}), $T_{a}(u)$ is the transfer matrix in the $a$-th antisymmetric irrep and $D_a$ stands for the bound state of $a$ derivatives. The main difference with the case encountered earlier is that the argument of the transfer matrix has been crossed from $u^{-\gamma}$ to $u^{\gamma}$. This small modification has an important effect, since in this crossed kinematics the transfer matrix is now of order $O(1)$ at weak coupling. This is easily verified to leading order at weak coupling, where the integrand can easily be expressed in terms of the Baxter polynomial $Q(u) = \prod_i (u-u_i)$ as
\beq\label{Q-form}
\frac{\textrm{int}^{3\gamma}_{a}}{\mathcal{A}} = \frac{Q(u^{[a+1]})+Q(u^{[-a-1]})-Q(u^{[a-1]})-Q(u^{[1-a]})}{Q(\frac{i}{2})}\, ,
\eeq
with $u^{[k]} = u+ik/2$. Finite size corrections in this opposed (i.e.~$\ell=\ell_{23}$) mirror channel are thus as big as they can be and start at $\ell+1$ loops. See figure~\ref{OPEConfs} for a summary.

\begin{figure}[t]
\begin{center}
\includegraphics[scale=0.31]{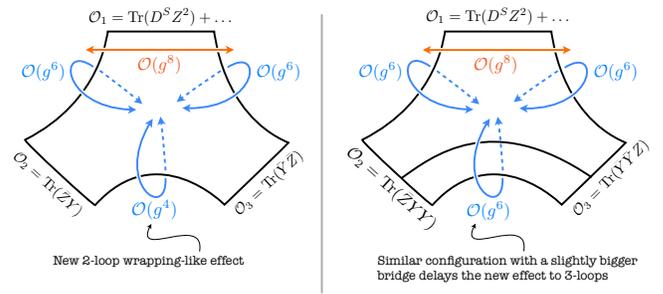}
\end{center}
\vspace{-3cm}
\caption{\normalsize The usual wrapping indicated at the top is delayed to four loops thanks to supersymmetry  while naively, based on length considerations alone, one would guess two loops. Then we have the new sort of finite size corrections coming from the exchange of virtual excitations between the two hexagon twist operators across each of the three bridges. 
The ones associated to the bridges adjacent to the non-protected operator are also delayed by supersymmetry but only to three loops. The one across it is not protected by SUSY and thus kicks in as early as at two loops, as indicated in the left. By slightly increasing the size of the bridge (i.e. by slightly increasing the twist of the BPS operators) we can delay this effect. On the right, for example, the bridge is increased by one unit delaying this new effect by one loop. 
} \la{OPEConfs} \vspace{-.2cm} 
\end{figure}


The next step is to perform the integration of the integrand above, as in~(\ref{calAwrapping}). In this regards it helps first massaging a bit the integral. The main trick is to bring the various terms in~(\ref{Q-form}) to the same level, by shifting their arguments appropriately. It yields the equivalent representation
\beq\label{upshot}
\frac{\delta\mathcal{A}_{{23}}}{\mathcal{A}} =\int \frac{du}{2\pi}  \hat{\mu}(u)\frac{Q(u)}{Q(\frac{i}{2})} + R + O(g^{2\ell_{23}+4})\, ,
\eeq
where $\hat{\mu}$ is an effective measure (with $\ell = \ell_{23}$ below)
\beq\label{eff-mu}
\begin{aligned}
\hat{\mu}(u) = &\sum_{a\geqslant 1}\bigg[\frac{\mu_{a}^{\gamma}(u^{[a+1]})}{(x^{[2a+1]}x^{+})^{\ell}} + \frac{\mu_{a}^{\gamma}(u^{[-a-1]})}{(x^{-}x^{[-2a-1]})^{\ell}} \\
&\qquad -\frac{\mu_{a}^{\gamma}(u^{[a-1]})}{(x^{[2a-1]}x^{-})^{\ell}} - \frac{\mu_{a}^{\gamma}(u^{[1-a]})}{(x^{+}x^{[1-2a]})^{\ell}}\bigg]\, ,
\end{aligned}
\eeq
and where $R = r_{+}+r_{-}$ is the sum of the residues at $u = \pm ia/2$, extracted along the way, 
\beq
r_{\pm} = \mp i\sum\limits_{a\geqslant 1}\underset{u=0}{\operatorname{{\rm res}}}\,\, \frac{\mu_{a}^{\gamma}(u^{[\mp a]})}{(x x^{[\mp 2a]})^{\ell}} \frac{Q(u\pm \frac{i}{2})}{Q(\frac{i}{2})}\, .
\eeq
This contour manipulation is perfectly well-defined for large enough bridge $\ell$, since then each term under the integral is separately integrable. What is not obvious, however, is that the resulting integral~(\ref{upshot}) remains convergent if, at the end, the number of magnon, i.e. the power of $Q$, exceeds the size of the bridge; which might happen quite early for many cases of interest. Remarkably enough, this is not the case, thanks to the sum over the bound states in~(\ref{eff-mu}) which endows the effective measure with an exponentially suppressed behaviour at large rapidity. The upshot is that the finite-size integral~(\ref{upshot}) in the $\ell=\ell_{23}$ channel is perfectly well-defined for any state, namely any polynomial $Q$.

For later use as well as for illustration, we shall conclude by presenting explicit expressions corresponding to the smallest possible bridge $\ell=\ell_{23}=1$ expanded to leading order at weak coupling. In this case, the local piece is given by
\beq\label{I}
R = g^4\frac{4\pi^4 q^{(0)}(\frac{i}{2}) +180i \zeta_3 q^{(1)}(\frac{i}{2})-5\pi^2 q^{(2)}(\frac{i}{2})}{60} 
\eeq
with $q^{(n)}(u) \equiv \partial^{n}_{u}(Q(u)+Q(-u))/Q(\frac{i}{2})$, and the integral part is controlled by the effective measure
\beq\label{J}
\hat{\mu}(u) = -(4\pi)^2 g^4\frac{1-12u^2+2\pi u(1+4u^2)\tanh{(\pi u)}}{(1+4u^2)^3\cosh^2{(\pi u)}}\, .
\eeq
It makes manifest the drastic improvement brought by the summation over bound states alluded to before, which here has led to the factor $\cosh^2(\pi u)$ in the denominator. The same feature can also be observed for higher values of $\ell=\ell_{23}$. At the level of the local piece $R$ one can see that increasing the bridge produces expressions of the type~(\ref{I}) but with higher derivatives of $Q$ and zeta values $\zeta_{z} = \zeta(z)$ of higher transcendentality.

It is now totally straightforward to compute the finite size effect by evaluating~(\ref{upshot}) with the appropriate wave function $Q$. This is what is done in the next section for the smallest unprotected 3-point functions.

\section{Advanced Weak Coupling Checks}

We can put both the asymptotic formula and its finite-size corrections to the test by comparing their predictions with perturbative weak coupling results. Computing directly three-point correlation functions of non-BPS in perturbation theory is a tremendous task, so much that beyond some notable exceptions \cite{oneLoopC123,1looptwist2,ThiagoJoao}, there are virtually no explicit results beyond one loop. (In view of all the conjectures put forward here it would be very interesting to try to improve this state of affairs.) 

Fortunately we can also read off structure constants -- or rather sum rules for products of two structure constants -- by analysing the operator product expansion of four points correlation functions. If the four operators are BPS, an impressive technology has been developed over the years for studying such correlators, see e.g. \cite{bigC,Grisha}. In the fusion of two BPS operators we get non-BPS operators and thus can access the structure constants $C^{\bullet\circ\circ}_{123}$ though the OPE of such correlators. The comparisons performed in this section will be the two-loop counterparts of the one loop checks done in \cite{tailoringNC}. We will present here three examples -- schematized in figure \ref{3examples} -- which illustrate different interesting features of these comparisons. 

\begin{figure}[t]
\begin{center}
\includegraphics[scale=0.31]{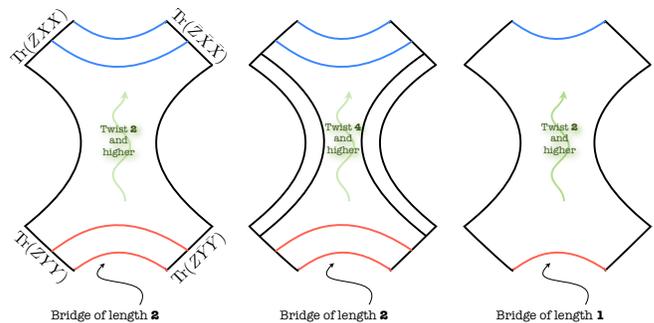}
\end{center}
\vspace{-2.5cm}
\caption{\normalsize By tuning the size and R-charge assignments of the external operators we probe different processes. The results \cite{bigC} for external operators of twist $3$ and $4$, for instance, allow us to consider three-point functions between two such BPS operators and one non-protected operator of twist $2$ or $4$ respectively, see left and middle pictures. We can do so while keeping the bottom bridge large enough as to suppress finite size corrections. From the results for smaller correlators as in \cite{Grisha}, we can extract similar perturbative data where finite size effects are now more relevant, see right picture. (See also figure~\ref{OPEConfs}.) 
} \la{3examples} \vspace{-.2cm} 
\end{figure}

The first example comes from analysing a correlation function of four BPS operators of size three as indicated on the left of figure \ref{3examples}. Such correlators were computed up to two loops in \cite{bigC}. By carefully choosing the R-charge of the external operators we can make sure to probe twist two operators (and higher) in the middle. This case is especially interesting since, for any spin $S$, there is a single twist two operator that contributes. As such, by analyzing the OPE of such four-point function we read off the corresponding structure constant without any ambiguity. Finally, starting with configurations with external operators of size three we have a bridge of size two opposite to the non-BPS operator which is flowing, see figure \ref{3examples}. As such, the associated finite size corrections are delayed to three loops, see figure \ref{OPEConfs}. The result is summarized in the following table:
\beq
\!\!\!\! \begin{array}{c|l}
S & \(\frac{C_{123}^{\bullet \circ \circ}}{C_{123}^{\circ \circ \circ}}\)^2 \text{ for twist $L=2$, bridge $\ell =2$ and spin $S$} 
\\ \hline
2 & \frac{1}{6}-2 g^2+28 g^4 +\dots \color{white}{\Big(} \\
4 & \frac{1}{70}-\frac{205 g^2}{882}+\frac{36653 g^4}{9261}+\dots  \color{white}{\Big(}\\
6 & \frac{1}{924}-\frac{553
   g^2}{27225}+\frac{826643623 g^4}{2156220000} +\dots  \color{white}{\Big(}\\
8 & \frac{1}{12870}-\frac{14380057
   g^2}{9018009000}+\frac{2748342985341731 g^4}{85305405235050000} +\dots  \color{white}{\Big(}\\
10 & \frac{1}{184756} \!-\!\frac{3313402433
   g^2}{27991929747600}\!+\!\frac{156422034186391633909
   g^4}{62201169404983234080000}+\dots  \color{white}{\Big(}
\end{array} \la{t1}
\eeq
We can bring our prediction to test and see if we can reproduce these intimidating numbers. As mentioned above, it suffices in this case to solve Bethe equations (\ref{BAE}) perturbatively for $L=2$ and plug the (unique) solution to these equations into our asymptotic all loop result (\ref{rank1}). In practice we do this semi-analytically by first solving Bethe equations numerically with a huge accuracy, plugging the solution into (\ref{rank1}) and rationalizing the end result. When the dust settles we perfectly reproduce all the values in table~(\ref{t1}).  

The second example is represented in the middle of figure \ref{3examples}. The main difference as compared to the previous example is that the operators flowing in the OPE now have at least twist four. The corresponding four-point correlator was also computed in \cite{bigC}. At a given spin there are now several operators, corresponding to the various solutions to Bethe equations for $L=4$. What we extract out of the OPE is now a prediction for a sum rule over all such solutions: 
\beq
\!\!\!\!\begin{array}{c|l}
S & \sum\(\frac{C_{123}^{\bullet \circ \circ}}{C_{123}^{\circ \circ \circ}}\)^2 \text{ for $L=4$, half-way split, and $\ell =2$}  \\ \hline 
 2 & \frac{7}{20}-2 g^2+14 g^4+ \dots  \color{white}{\Big(}\\
 4 & \frac{4}{63}-\frac{49 g^2}{81}+\frac{12073
   g^4}{1944}+\dots \color{white}{\Big(} \\
 6 & \frac{29}{3432}-\frac{468382 g^2}{4601025}+\frac{23583608243837
   g^4}{18948861360000}+\dots \color{white}{\Big(} \\
 8 & \frac{23}{24310}-\frac{12573551
   g^2}{957283600}+\frac{20041636847534339
   g^4}{111957380327520000}+\dots  \color{white}{\Big(}\\ 
   10 & \frac{67}{705432}-\frac{1686168467
   g^2}{1159429042800}+\frac{33947407541075800567
   g^4}{1585464490978179840000}+\dots  \color{white}{\Big(}
\end{array}
\eeq
Since the bridge is still of size $\ell=2$ we can still test these predictions against our asymptotic three-point predictions and find a perfect match. (Such kind of comparisons were done in great detail in \cite{tailoringNC} at one loop level and it is straightforward to simply expand the asymptotic result one loop further.)    

The final example, represented in the right of figure \ref{3examples}, is extracted from the correlator with four external operators of size two. These belong to the famous family of correlators of stress-tensor supermultiplets for which amazing progress has been achieved very recently \cite{Grisha}. These are by far the most well studied of all BPS four-point correlation functions and the only one that has been computed and OPE decomposed beyond two loops \cite{Grisha}. 
Here, we shall restrict our analysis to two loops. 

For us this example is particularly interesting since it is the simplest example containing a bridge of minimal size $\ell=1$ facing the non-BPS operator. As discussed in the previous section this is expected to lead to a new sort of finite size effect computed in (\ref{upshot}). From the OPE point of view, the extraction of the structure constant is as simple as for the first example and yields 
\beq
\!\!\!\! \begin{array}{c|l}
S & \(\frac{C_{123}^{\bullet \circ \circ}}{C_{123}^{\circ \circ \circ}}\)^2 \text{ for twist $L=2$, bridge $\ell =1$ and spin $S$}
\\ \hline   
2 & \frac{1}{6}-2 g^2+(28+12 \zeta (3)) g^4+\dots \color{white}{\Big(} \\
 4 & \frac{1}{70}-\frac{205
   g^2}{882}+\left(\frac{76393}{18522}+\frac{10 \zeta (3)}{7}\right)
   g^4+\dots \color{white}{\Big(} \\
 6 & \frac{1}{924}-\frac{553
   g^2}{27225}+\left(\frac{880821373}{2156220000}+\frac{7 \zeta
   (3)}{55}\right) g^4+\dots \color{white}{\Big(} \\
 8 & \frac{1}{12870}\!-\!\frac{14380057
   g^2}{9018009000}+\!(\!\frac{5944825782678337}{170610810470100000}\!+\!\frac{761 \zeta (3)}{75075}) g^4+\dots \color{white}{\Big(}\!\!\!\!\!\!\!\!\!\!\!\! \\
 10 & \frac{1}{184756}-\frac{3313402433
   g^2}{27991929747600} \\ 
&  \qquad  +\left(\frac{171050793565932326659}{622011694049
   83234080000}+\frac{671 \zeta (3)}{881790}\right)
  g^4+ \dots
  \color{white}{\Big(}
   \end{array}\label{t3}
\eeq
We readily note that at tree level and one loop the values obtained here are exactly the same as in the first example, see table (\ref{t1}). This is to be expected since at tree level and one loop both examples can be computed through the asymptotic result which is insensitive to the bridge size. At two loops, however, we now know that we should correct the asymptotic result by the new finite size correction (\ref{upshot}). The latter is only sensitive to the leading order Bethe roots, which for twist two are given by the roots of Hahn polynomials \cite{Hahn} 
\beq
Q(u) \propto  \,_3 F_2\left(-S,S+1,\frac{1}{2}- iu ;1,1;1\right)  \,. 
\eeq
Plugging these polynomials in (\ref{I}) and (\ref{J}) and adding up these contributions we perfectly reproduce the missing contribution thus obtaining precisely the values in table~(\ref{t3})! This  is definitely the most non-trivial check of our construction thus far. It would be very interesting to extend this comparison one loop further given the existing data.

It would also be very interesting to obtain higher-loop results for four-point correlators of BPS operators of arbitrary size. As illustrated above, by properly playing with the external R-charge assignments of the four operators we can probe different effects. By having large families of such correlators we would be able to isolate the asymptotic result from the various finite size corrections thus greatly simplifying the comparison with the integrability conjectures. It would be very interesting, for instance, to probe the effect of the dressing phase in the asymptotic result without the contaminations by complicated finite size effects; this would be straightforward were three-loop results for such correlators available. In the meantime, in the absence of such results, we can use strong coupling as a clean probe of this phase; this is what we turn to in the next section.

\begin{figure}[t]
\begin{center}
\includegraphics[scale=0.3]{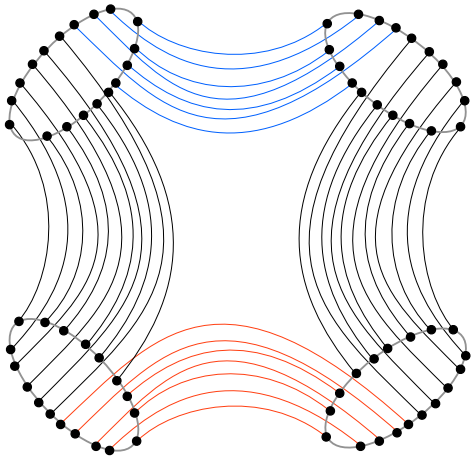}
\end{center}
\caption{\normalsize With two and three-point functions at hand, one can in principle reconstruct any correlation function. The simplest correlators to attack are probably those involving large operators, connected by large bridges. For those, the contribution of double traces is suppressed and the structure constants arising in their OPE decomposition are simply given by their asymptotic expressions.
} \la{Big4} \vspace{-.2cm} 
\end{figure}

Conversely, it would be fascinating to try to glue together the three-point functions proposed herein into higher-point correlators. As a first step we could envisage studying large operators, with large bridges of propagators connecting any two consecutive operators as in a painting with a large frame. Such correlators, which one would naturally dub as \textit{asymptotic four point functions} (see figure \ref{Big4}) should be sensitive to the asymptotic part of the structure constants up to very high loop order and, non unrelated, will have mixing with double traces delayed as well. They should be considerably simpler than their finite size counterparts which we have been analyzing here and would probably be the ideal playground for learning how to stitch our pair of pants into more complicated correlators. For the spectrum problem -- and for the three-point proposal put forward here -- the detour into the world of operators clearly pays off. It would be only natural that history would repeat itself for four-point correlators. For interesting works where such large correlators are considered see \cite{4ptWeak} at weak coupling and \cite{4ptStrong} at strong coupling. 


\section{Strong Coupling}

We will now study our proposals at strong coupling. The strong coupling regime is a very instructive laboratory for probing both the asymptotic result as well as its finite size corrections. (This was already the case for the spectrum problem.) The underlying reason is that in this relativistic regime, physical and mirror particles behave very symmetrically. 

The strong coupling result is known \cite{Janik:2011bd,shotasu2,Kazama:2013qsa,shota2}. Here we will show that our proposal has the right features to reproduce it as we send the coupling to infinity. 

To start with, let us consider the asymptotic contribution (\ref{rank1}) (for two BPS and one non-BPS operator). To match with the classical computation we take the standard classical limit corresponding to a large number of magnons with large rapidities $u_j ,S, L \sim \sqrt{\lambda}$. In this limit, the sum over partitions in (\ref{rank1}) can be replaced by a path integral over the densities $\rho_\alpha$ and $\rho_{\bar \alpha}$, which in turn can be evaluated by a saddle point computation following \cite{tailoring3}. The reason why we can straightforwardly recycle that weak coupling computation is that the absolute value of $f$ -- which governs the singular behaviour for nearby roots -- is the same as at weak coupling while the phase of $f$ will lead to the simple replacement of the weak coupling quasi-momenta by its strong coupling counterpart: 
\beq
f(u_i,u_j)^2 = \underbrace{f(u_i,u_j)f(u_j,u_i) }_{\simeq {\frac{(u_i-u_j)^2+1}{(u_j-u_j)^2}}+O(1/g^2)} \times  \, \underbrace{{\frac{f(u_i,u_j)}{f(u_j,u_i)}} }_{S(u_j,u_i)} \, . \nn
\eeq
As such, the computation of the so-called stochastic anomaly is the same as at weak coupling and we can simply plug the strong coupling quasi-momenta \cite{KMMZ} in the final result (3.28) in \cite{tailoring3}. The computation of the norm also follows the weak coupling computation in a straightforward fashion. In short, we see that our asymptotic contribution nicely exponentiates into 
\beq
\left. C_{123}^{\bullet\circ\circ} \right|_\text{asymptotic} \!= e^{ \oint \! \frac{du}{2\pi} \[ \text{Li}_2(e^{i  p_1+i  p_2-i    p_3})-\frac{1}{2} \text{Li}_2(e^{2i  p_1}) \]}  \la{strongA}
\eeq
where the contour should encircle the cuts of the corresponding classical algebraic curve. This matches with the string theory result~\cite{Kazama:2013qsa} including the contour~\cite{shota2}. In (\ref{strongA}) the quasi-momenta should be understood as the sphere quasi-momenta $\tilde p_i(x)$ for the $SU(2)$ case and as their AdS counterparts $\hat p_i(x)$ for the $SL(2)$ case. In either case, the quasi-momenta $p_2$ and $p_3$ are those of the BMN string, i.e. $p_2= 2\pi x L_2/((x^2-1)\sqrt{\lambda})$ and similarly for~$p_3$. 

Next we can include the effects of mirror excitations. Including them all is beyond the scope of this paper. Instead we will see how the leading effect corresponding to a single virtual excitation matches the leading classical finite size correction. For simplicity we consider the case worked out above corresponding to a mirror excitation in the opposed mirror channel, see (\ref{voila1}). As at weak coupling, we will consider the non-BPS operator to be of $SL(2)$ type. One can readily take the strong coupling classical limit of  (\ref{voila1}), see  (\ref{Tp}). In the end, the result can be summarized as correcting the exponent in~(\ref{strongA}) by
\beq\label{mirror-strong}
\begin{aligned}
\int\limits_{U^{-}} \frac{du}{2\pi} e^{i   p_{2}(x)+ i   p_{3}(x)}\big[e^{ -i \hat p_{1}(x)} + e^{i  \hat p_{1}(1/x))}  -2 e^{-i \tilde p_{1}(x)} \big]
\end{aligned}
\eeq
where $\hat p_1(x)$ and $\tilde p_1(x) = 2\pi x L_1/((x^2-1)\sqrt{\lambda})$ are the AdS and Sphere quasi-momenta mentioned earlier.
Its four-term structure is directly inherited from the form of the transfer matrix, with the first two terms being interpreted as originating from $AdS$ and the last two from the sphere in the dual string description, see appendix~\ref{tm}. Last but not least, an important ingredient in this expression is the contour of integration $U^{-}$, which runs from $x=-1$ to $x=+1$ along the lower half unit circle, as it should be for a mirror (perturbative) magnon.

Remarkably, the formula~(\ref{mirror-strong}) matches perfectly with the string theory prediction~\cite{shota2}, see appendix \ref{strongAp}. Let us stress again that this expression only captures the leading wrapping correction along the bridge $\ell_{23}$. The string theory result contains all these single bridge effects as well as all the more complicated processes with (arbitrarily many) particles navigating through all the mirror channels of the 3-point function. It will be fascinating to verify that the complete finite-size structure of the string answer can be reproduced from the hexagon approach. 

Finally, in \cite{ivan}, powerful techniques were proposed to tackle classical limits and their leading semi-classical correction at weak coupling. It would be very interesting to investigate how much can be borrowed from this technology to tackle strong coupling loop effects beyond the classical area result.

\section{Conclusion}

\begin{figure}[t]
\begin{center}
\includegraphics[scale=0.31]{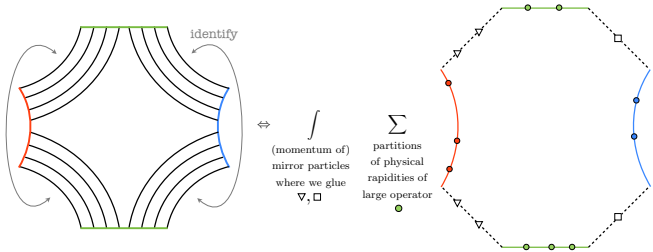}
\end{center}
\vspace{-3.5cm}
\caption{\normalsize 
The collision of two hexagons in a non-extremal process amounts to removing one of the bridges between two operators. This leads to a picture where extremal correlators can be described by an octagon twist operator of sort. This can also be directly depicted at weak coupling, as illustrated on the left. In this case we cut the correlator only twice and, as such, we get a single sum over partitions of the roots of the largest operator. See appendices \ref{ap-a} and \ref{ap-b} for more details. 
} \la{octagon} \vspace{-.2cm} 
\end{figure}

In this letter, we presented a new framework for computing three-point functions of single trace operators in planar $\mathcal{N}=4$ SYM theory. The approach
introduces a new elementary object, called hexagon operator, whose form factors, when properly glued together, should determine the three-point functions at any value of the coupling and for arbitrary operators. 

In short, we have argued that the hexagon form factors fulfill the same axioms as the $PSU(2|2)$ invariant S-matrix constructed by Beisert, if not for unitarity. Beyond two particles, our solution stands as a conjecture since it implicitly makes assumption of factorization \`a la Yang Baxter, which is not a priori a requirement for form factors. This ansatz is consistent with the Watson equation and the decoupling condition and it would be nice to see if both requirements can actually imply the factorization of the form factor. This could provide us with a more complete or first-principle derivation of our ansatz. Along these lines it would be nice to investigate if the Yangian symmetry can be added to the bootstrap laid out here.

Our final proposal identifies the structure constant with a finite volume correlator of two hexagon operators. We further proposed a finite coupling IR form factor expansion for such correlator which is most tractable as long as the two hexagons are well separated. In the language of three-point functions, this corresponds to being far from extremal. An equally interesting limit to explore would be the UV regime of sort where the two hexagons are now almost on top of each other. This should be relevant for the study of extremal correlators and, likely, to closely connected problems concerning the mixing with double traces, $1/N$ correction to anomalous dimensions and so on. What our picture suggests is that the collision of two hexagons should be governed by an octagon operator, see figure \ref{octagon} and further discussion in appendices \ref{ap-a} and \ref{ap-b}. 
It would be very interesting to develop the octagon bootstrap in detail.

There are several other interesting directions worth pursuing, such as continuing the weak coupling comparisons to higher loops and carrying out the full strong coupling analysis for the classical exponent and beyond. It would also be fascinating to make contact with the all-loop predictions at large spin ~\cite{Alday:2013cwa} or at the Regge pole ~\cite{miguel},
to add boundaries to our problem such as to describe correlators of open spin chains inserted along BPS Wilson lines and perhaps also some mixed open/closed-string correlators. 
It would also be nice to make connection with alternative approaches describing quantum eigenstates of integrable models. In particular, trying to cast our results in the language of Sklyanin separation of variables \cite{Sklyanin} might pave the way towards the embedding of the three-point correlator story within the extremely powerful quantum spectral curve framework \cite{Gromov:2013pga}.



%

Finally, it would be fascinating to look for a relationship between spin-chain hexagon transitions and null Wilson loop pentagon transitions. This could lead to a more unified picture of the elementary worldsheet patches underpinning the correlators of the gauge/string theory.

\section*{Acknowledgements} We are obliged to N.~Berkovits, D.~Gaiotto,  N.~Gromov, V.~Kazakov, I.~Kostov, R.~Janik, J.~Maldacena, D.~Serban and A.~Sever for illuminating discussions. We are specially thankful to N.~Gromov and A.~Sever for the stimulating collaborations and brainstorming sessions on the problem addressed in this paper and closely related ones. {We are grateful to D.~Serban, A.~Sever, J.~Maldacena, and in particular T.~Fleury for helpful comments on the manuscript.} We also acknowledge all the participants of the `Program on Integrability, Holography and the Conformal Bootstrap' for an inspiring and exciting program. We would like to thank FAPESP grant 2011/11973-4 for funding our visit from Nov 2014 to March 2015 to ICTP-SAIFR where most of this work was done.
Research at the Perimeter Institute is supported in part by the Government of Canada through NSERC and by the Province of Ontario through MRI. 

\appendix 

\section{Hexagon twist operator from the worldsheet\label{ap-a}}

In this appendix, we show that the hexagon we studied in the main text can be understood as a twist operator creating a conical excess $\pi$ on the string worldsheet. At strong coupling $\lambda\gg1$, the three-point function of long single-trace operators admits an alternative description in terms of the classical string worldsheet in AdS. When the operators are scalar, this problem was studied in \cite{Janik:2011bd,Kazama:2013qsa} using the Pohlmeyer description of the minimal surface. In that approach, the induced metric of the surface is expressed as
\beq
ds^2 = \sqrt{T\bar{T}} \cosh \gamma \,dz d\bar{z}
\eeq
where $T$ and $\bar{T}$ are the AdS part of the worldsheet stress-energy tensor, which behave near the insertion point of the operator as
\beq
\begin{aligned}
T(z) \sim \frac{L_k^2/\lambda}{z-z_k}\quad (z\to z_k) \,.
 \end{aligned}
 \eeq
Here $L_k$ is the length (the R-charge) of the $k$-th vertex operator. The function $\gamma$ is the solution to the modified sinh-Gordon equation and is in general a complicated function of the worldsheet coordinates. However, when the operators are $1/2$ BPS and the lengths are all large ($L_k\gg \sqrt{\lambda}$), $\gamma$ vanishes almost everywhere and the induced metric approaches
\beq
ds^2 \simeq \sqrt{T\bar{T}}\,dz \,d\bar{z}.\label{induced}
\eeq
In the vicinity of a (simple) zero of $T(z)$, this metric takes the form
\beq
ds^2 \propto r\,(dr^2 +r^2 d\theta^2) \quad (r\sim 0)\,,
\eeq
which is precisely the one describing the conical singularity with the excess angle $\pi$. In the case of three-point functions, there are generically two such zeros, whose distance roughly scales as the length of the bridges $(L_i+L_j-L_k)/2$. This corroborates our interpretation of the three-point function as the two-point function of hexagon twist operators.

As we decrease the lengths of a bridge, these two zeros come close to each other and eventually collide when the bridge disappears, leaving a point with the excess angle $2\pi$. Geometrically, this can be seen as a process in which two hexagons merge and form one octagon. This is indeed the case for extremal three-point functions, where the length of the third operator is given by the sum of the lengths of the other two. Interestingly, the same conical singularity also appears at the interaction points in the Mandelstam diagram  \cite{Mandelstam} of the lightcone-gauge string theory. These observations suggest that the octagon twist operator might provide a good description for the extremal three-point functions and the spltting-and-joining processes in the lightcone gauge. In appendix \ref{ap-b}, we will see explicitly that the pp-wave string field theory vertex exhibits the properties of the octagon twist operator.

\section{Twisted translation and symmetry} \la{symmetryAp}
In the integrability-based approach to $\mathcal{N}=4$ SYM, we first choose a particular observable, which plays the role of the ``vacuum'' of the spin chain, and then describe other observables by putting ``excitations'' on top of it. For the spectrum problem, the vacuum is provided by the two-point function of $1/2$ BPS operators 
\beq
\langle \tr Z^{L}(0)\,\, \tr \bar{Z}^{L} (\infty) \rangle\,.\label{twopnt}
\eeq
Upon this choice of the vacuum, the $PSU(2,2|4)$ symmetry of $\mathcal{N}=4$ SYM is broken down to $PSU(2|2)^2$ generated by (the notation follows the one in \cite{Beisert}.)
\beq
\begin{aligned}
&L^{\alpha}{}_{\beta}, \quad \dot{L}^{\dot{\alpha}}{}_{\dot{\beta}}, \quad R^{a}{}_{b},\quad \dot{R}^{\dot{a}}{}_{\dot{b}} ,\\
&Q^{\alpha}{}_{b}, \quad \dot{Q}^{\dot{\alpha}}{}_{\dot{b}},\quad S^{a}{}_{\beta},\quad \dot{S}^{\dot{a}}{}_{\dot{\alpha}},
\end{aligned}\label{psu2}
 \eeq
and the excitations forms the multiplet of this symmetry.

For three-point functions, an appropriate ``vacuum'' configuration is provided by the three-point function of $1/2$-BPS operators
\beq
\mathcal{O}_i = \tr(\vec{Y}_i\cdot \Phi)^{L_i}\left(x_i^{\mu}\right)\, ,
\eeq
where $\vec{Y}_i$'s are six-component complex vectors satisfying $\vec{Y}_i\cdot \vec{Y}_i=0$. To study the symmetry, it is convenient to put three operators on a line by performing a suitable conformal transformation. Similarly, using the R-symmetry transformation, we can align $\vec{Y}_i$'s along a particular $U(1)$ direction inside $SO(6)$. Combining these two transformations, we can always bring the three operators to 
\beq
\langle \tr \mathfrak{Z}^{L_1}(a_1)\, \tr \mathfrak{Z}^{L_2}(a_2)\,\tr \mathfrak{Z}^{L_3}(a_3)\rangle\label{tt3pt}
\eeq
where $\mathfrak{Z}$ is the ``twisted-translated'' scalar \cite{DrukkerPlefka} defined by
\beq
\begin{aligned}
\mathfrak{Z}(a) &\equiv e^{\mathcal{T}a}\cdot Z(0) e^{-\mathcal{T}a}\\
&=\left(Z+ \mm^2 a^2 \bar Z +\mm a (Y-\bar{Y})\right) (0,a,0,0)\,
\end{aligned}
\eeq
with
\beq
\mathcal{T}\equiv -i\epsilon_{\alpha\dot{\alpha}} P^{\dot{\alpha}\alpha}+\mm\epsilon_{\dot{a}a}R^{a\dot{a}}\,.
\eeq
Here, $\mm$ is a quantity with mass dimension $1$, which can be set to unity by performing a further dilatation.

Among the $PSU(2|2)^2$ generators (\ref{psu2}), the ones which commute with the twisted translation $\mathcal{T}$ are given by
\beq
\begin{aligned}
&\mathcal{L}^{a}{}_{b}=L^{\alpha}{}_{\beta}+\dot{L}^{\dot{\alpha}}{}_{\dot{\beta}},&&\mathcal{Q}^{\alpha}{}_{a}=Q^{\alpha}{}_{a}+i\mm \epsilon^{\alpha\dot{\beta}}\epsilon_{a\dot{b}}\dot{S}^{\dot{b}}{}_{\dot{\beta}},\\
&\mathcal{R}^{a}{}_{b}=R^{a}{}_{b} + \dot{R}^{\dot{a}}{}_{\dot{b}}, &&\mathcal{S}^{a}{}_{\alpha}=S^{a}{}_{\alpha} + \frac{i}{\mm}\epsilon^{a\dot{b}}\epsilon_{\alpha\dot{\beta}} \dot{Q}^{\dot{\beta}}_{\dot{b}}.
\end{aligned}\label{diagonal}
\eeq
(The part generated by $\mathcal{L}^{a}{}_{b}$ is the $O(3)$ rotation around the line on which the operators are inserted.)
Thus, the hexagon vertex (in the absence of excitations) preserves the diagonal $PSU(2|2)$ symmetry generated by (\ref{diagonal}).

Non-BPS three-point functions are obtained by putting excitations on top of (\ref{tt3pt}). More precisely, we first consider three non-BPS operators constructed on the $Z$-vacuum at $x^{\mu}=0$ and then perform the twisted translation:
\beq
\begin{aligned}
&\mathcal{O}_i =  \left.\tr \left( Z X ZZ\cdots\right)+\cdots\right|_{x^{\mu}=0}\\
&\overset{\mathcal{T}}{\to} \mathfrak{O}_i(a_i)= e^{\mathcal{T}a_i}\cdot\mathcal{O}_i\cdot e^{-\mathcal{T}a_i} 
\end{aligned}
\eeq
When the operators $\mathcal{O}_i$ satisfy the highest weight conditions $\epsilon^{\dot{a}b}R_{\dot{a}b} \cdot \mathcal{O}_i=0$ and $\epsilon^{\alpha\dot{\alpha}}K_{\alpha\dot{\alpha}}\cdot~\mathcal{O}_i=0$, as is the case for on-shell Bethe states with finite rapidities, the $a_i$-dependence is determined \cite{Kazama:2014sxa} by the Ward identity as
\beq
\begin{aligned}
&\langle\mathfrak{O}_1(a_1)\mathfrak{O}_2(a_2)\mathfrak{O}_3(a_3) \rangle\\
& = \frac{ \mm^{J_1+J_2+J_3} \mathbb{C}_{123}}{(a_1-a_2)^{\delta_{12|3}}(a_2-a_3)^{\delta_{23|1}}(a_3-a_1)^{\delta_{31|2}}}\,.
\end{aligned}
\eeq
Here $J_i$ is the R-charge of the operator $\mathcal{O}_i$ and $\delta_{ij|k}$ is given by
\beq
\delta_{ij|k}=(\Delta_i+\Delta_j-\Delta_k)-(J_i+J_j-J_k)
 \eeq 
The hexagon vertex computes $\mathbb{C}_{123}$, which is the part independent of $a_i$'s. Note that this may not coincide with the usual definition of the structure constant since $\mathbb{C}_{123}$ can also contain the tensor structures if $\mathcal{O}_i$'s have the transverse excitations.

\section{Symmetry and form factors}\label{sym-ff}

In this appendix we give more details on the construction of the hexagon form factors for 1 and 2 particles.

To begin with, we recall that each magnon transforms in the bifundamental representation of  $PSU(2|2)^2 \ltimes \mathbb{R}^3$, where the three central charges $C, P, K$ are shared among the left and right extended $\mathfrak{psu}(2|2)$ subalgebras~\cite{Beisert:2006qh}. Out of all these generators, only the linear combinations spanning the diagonal $\mathfrak{psu}(2|2)$ subalgebra annihilate the vertex. The corresponding $6|8$ (super)generators are spelled out in~(\ref{diagonal}). In fact, the closure of the diagonal algebra requires including the diagonal central charge
\beq\label{d-central}
\mathcal{P} = P-\kappa^2 K\, ,
\eeq
which appears in the anticommutators $\{\mathcal{Q}, \mathcal{Q}\} , \{\mathcal{S}, \mathcal{S}\} \sim \mathcal{P}$. (Note that this is the only central charge extending the diagonal subalgebra; importantly, the energy $C$, whose conservation by the vertex would have led to trivial solution, does not appear.)  To enforce the diagonal symmetry, we should make sure that the central element~(\ref{d-central}) kills the hexagon vertex, i.e.~that
\beq\label{P-constraint}
\left<\mathfrak{h}|\mathcal{P}|\psi\right> = 0
\eeq
holds for a generic spin chain state $\psi = \chi_{1},\ldots , \chi_{n}$.

The implementation of the constraint~(\ref{P-constraint}) depends on how we address the mild non-locality of the chain of fundamentals used for describing the state, see~\cite{Beisert:2006qh}. Here we shall consider two possibilities we loosely refer to as spin chain frame and string frame. (The relation between the two frames will be discussed in appendix \ref{mapping}.) They correspond to the twisted notation of~\cite{Beisert:2006qh} and to the string conventions of~\cite{reviewHuge}, respectively. In both cases, the central charges $P, K$, and consequently $\mathcal{P}$, act diagonally on a generic state, but while this is literally true in the string picture of~\cite{reviewHuge}, this holds up to addition or deletion of a vacuum $Z$ unit in the twisted spin chain picture of~\cite{Beisert:2006qh}. In the latter case, enforcing~(\ref{P-constraint}) requires first understanding how to deal with the $Z$ marker at the level of the form factors. This is the case we will first consider.

In the spin chain picture, equation~(\ref{P-constraint}) is equivalent to
\beq\label{sc-constraint}
0 = g\alpha (1-e^{ip}) \big<\mathfrak{h}|Z^{+}\psi\big> - \frac{g \kappa^2}{\alpha} (1-e^{-ip}) \big<\mathfrak{h}|Z^{-}\psi\big>\, ,
\eeq
where $\alpha$ is a parameter common to the left and right representation (here and below, we follow the notations used in Section~3.5 of~\cite{Beisert:2006qh}, if not for the additional twist associated to the marker $\mathcal{Y}$ which shall be discarded.)
We solve this equation by assuming that the $Z$ marker is diagonalized by the vertex, such that $\big<\mathfrak{h}| Z^{n}\psi\big> = z^{n}\big<\mathfrak{h}|\psi\big>$ holds true. The $z$ eigenvalue can then be fixed, up to a sign, by means of equation~(\ref{sc-constraint}), giving
\beq\label{z-rule}
z^2 = -\frac{\kappa^2}{\alpha^2} e^{-ip}\, ,
\eeq
where $p$ stands for the total momentum of the state. Equipped with the rule~(\ref{z-rule}) for handling the $Z$ marker, it becomes extremely easy to derive SUSY Ward identities for the form factors.

Consider the case of a single magnon first. Then by $O(3)\times O(3)$ symmetry it must take the form~(\ref{1-point}) for some $\N$. By using now that $\big<\mathfrak{h}|$ is annihilated by the right action of the supercharges we can relate $\N$ to the parameters of the left and right irreps. E.g., imposing
\beq
0 = \big<\mathfrak{h}|\mathcal{Q}^{\alpha}_{\,\,\,\, a}\big|\Psi^{a\dot{\beta}}\big> \equiv \big<\mathfrak{h}|\mathcal{Q}^{\alpha}_{\,\,\,\, a}\big|\phi^{b}\psi^{\dot{\beta}}\big> 
\eeq
immediately gives
\beq\label{normal}
\N =  \frac{\kap (x^- - x^+)}{\gamma\dot{\gamma}}\, ,
\eeq
where $\gamma, \dot{\gamma}$ are free parameters associated to relative normalization between boson and fermion in the left/right multiplet, as in~\cite{Beisert:2006qh}. For unitary representations $|\gamma| = |\dot{\gamma}| = \sqrt{ix^{-}-ix^{+}}$ and thus $|N| = |\kappa|$. Clearly, by properly choosing the phases and the parameter $\kappa$, we can ensure that $N = 1$.

A similar analysis can be applied to two magnons. The $O(3)\times O(3)$ symmetry now requires that the form factor takes the general form
\beq
\begin{aligned}
\big<\V\big| \Phi^{a\dot{a}}_{1}\Phi^{b\dot{b}}_{2}\big> &= \mathcal{A}_{12} \epsilon^{a\dot{b}}\epsilon^{b\dot{a}} + \frac{1}{2}(\mathcal{A}_{12}-\mathcal{B}_{12}) \epsilon^{ab}\epsilon^{\dot{a}\dot{b}} \, , \\
\big<\V\big| \Phi^{a\dot{a}}_{1}\mathcal{D}^{\beta\dot{\beta}}_{2}\big> &= \mathcal{G}_{12}\epsilon^{a\dot{a}}\epsilon^{\beta\dot{\beta}}\, , \, \, \, \big<\V\big| \mathcal{D}^{\alpha\dot{\alpha}}_{1}\Phi^{b\dot{b}}_{2}\big> = \mathcal{L}_{12}\epsilon^{\alpha\dot{\alpha}}\epsilon^{b\dot{b}}\, , \\
\big<\V\big| \mathcal{D}^{\alpha\dot{\alpha}}_{1}\mathcal{D}^{\beta\dot{\beta}}_{2}\big> &= \mathcal{D}_{12}\epsilon^{\alpha\dot{\beta}}\epsilon^{\beta\dot{\alpha}} + \frac{1}{2}(\mathcal{D}_{12}-\mathcal{E}_{12})\epsilon^{\alpha\beta}\epsilon^{\dot{\alpha}\dot{\beta}}\, , \\
\big<\V\big| \Psi^{a\dot{\alpha}}_{1}\Psi^{b\dot{\beta}}_{2}\big> & = \frac{1}{2}\mathcal{C}_{12}\epsilon^{ab}\epsilon^{\dot{\alpha}\dot{\beta}}\, , \,\,\, \big<\V\big| \Psi^{a\dot{\alpha}}_{1}\Psi^{\beta\dot{b}}_{2}\big> = \mathcal{H}_{12}\epsilon^{a\dot{b}}\epsilon^{\beta\dot{\alpha}}\, , \\
\big<\V\big| \Psi^{\alpha\dot{a}}_{1}\Psi^{b\dot{\beta}}_{2}\big> &= \mathcal{K}_{12}\epsilon^{b\dot{a}}\epsilon^{\alpha\dot{\beta}}\, , \,\,\, \big<\V\big| \Psi^{\alpha\dot{a}}_{1}\Psi^{\beta\dot{b}}_{2}\big> = \frac{1}{2}\mathcal{F}_{12}\epsilon^{\dot{a}\dot{b}}\epsilon^{\alpha\beta}\, ,
\end{aligned} \la{all2pt}
\eeq
for some yet-to-be-determined coefficients $\mathcal{A}_{12}, \ldots , \mathcal{L}_{12}$. Imposing then that 
\beq\label{WI}
0 = \big<\mathfrak{h}|\mathcal{Q}^{a}_{\,\,\,\, \alpha}\big|\Phi^{b\dot{c}}_{1}, \Psi_{2}^{c\dot{\beta}}\big> = \big<\mathfrak{h}|\mathcal{S}^{\alpha}_{\,\,\,\, a}\big|\Phi^{b\dot{c}}_{1}, \Psi_{2}^{c\dot{\beta}}\big>\, ,
\eeq
gives four linear relations relating five of the above amplitudes, and similar identities would be obtained by plugging other states in~(\ref{WI}). In the end, the set of all such relations allows us to express the form factors as
\beq
\begin{aligned}\label{final-app}
\mathcal{A}_{12} &= h_{12}A_{12}\, , \qquad \mathcal{B}_{12} = h_{12}B_{12}\, , \\
\mathcal{G}_{12} &= h_{12}\N_{2}G_{12}\, , \qquad \mathcal{L}_{12} = h_{12}\N_{1}L_{12}\, , \\
\mathcal{D}_{12} &= -h_{12}\N_{1}\N_{2}D_{12} \, , \qquad \mathcal{E}_{12} = -h_{12}\N_{1}\N_{2}E_{12}\, ,\\
\mathcal{C}_{12} &= -h_{12}\N_{1}\N_{2}z^{-1} C_{12}\, , \qquad \mathcal{F}_{12} = -zh_{12}F_{12}\,, \\
\mathcal{K}_{12} &= h_{12}\N_{2}K_{12}\, , \qquad \mathcal{H}_{12} = -h_{12}\N_{1}H_{12}\, , 
\end{aligned}
\eeq
with $\N$ and $z$ as defined in~(\ref{normal}) and~(\ref{z-rule}), with $p = p_{1}+p_{2}$ in the latter case, and with the coefficients $A_{12}, \ldots , L_{12}$ as in the table 1 of~\cite{Beisert:2006qh} (with a trivial scalar factor, i.e.~with $S_{12}^{0} = 1$, and with $\xi_i =1$). 

For short, we just verified, by explicit algebra, that the diagonal $PSU(2|2)$ symmetry group forces the two-magnon form factor to be :
\beq\label{2pt}
\mathfrak{h}_{12}^{A\dot{A}, B\dot{B}} = (-1)^{\dot{f}_{1}f_{2}} S^{AB}_{CD}(1,2) \mathfrak{h}_{1}^{D\dot{A}}\mathfrak{h}_{2}^{C\dot{B}}\, ,
\eeq
where $f_i, \dot{f}_i$ is the left/right fermion number of the $i$-th magnon, and where we brought the S-matrix~\cite{Beisert:2006qh} to a matrix form,
\beq\label{Smf}
\mathcal{S}_{12}|\chi_{1}^{A}\chi_{2}^{B}\big> = S^{AB}_{CD}(1,2) \big|\chi_{2}^{C}\chi_{1}^{D}\big>\, ,
\eeq
using our rule~(\ref{z-rule}) for pulling out the $Z$ marker from the out-states in~\cite{Beisert:2006qh} whenever needed (this only concerns the $\phi\phi \leftrightarrow \psi\psi$ scattering amplitudes $C_{12}$ and $F_{12}$). As before, the scalar factor in~\cite{Beisert:2006qh} is assumed to be $1$ in~(\ref{Smf}). A fully equivalent way of writing it is as
\beq\label{2pt-bis}
\mathfrak{h}_{12}^{A\dot{A}, B\dot{B}} = (-1)^{\dot{f}_{1}f_{2}} \dot{S}^{\dot{A}\dot{B}}_{\dot{C}\dot{D}}(1,2) \mathfrak{h}_{1}^{\dot{D}A}\mathfrak{h}_{2}^{\dot{C}B}\, ,
\eeq
where $\dot{S}$ is the right S-matrix, obtained from the left one by dotting whatever can be dotted. Equations~(\ref{2pt}) and~(\ref{2pt-bis}) correspond to the two equivalent way of evaluating our general ansatz, obtained by scattering either the left or the right parts of the magnons, as depicted in figure~\ref{LRactio}.
\begin{figure}[t]
\begin{center}
\includegraphics[scale=0.31]{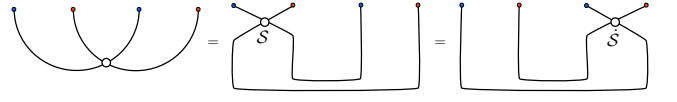}
\end{center}
\vspace{-5.4cm}
\caption{\normalsize The S-matrix in the form factor ansatz can be thought of as acting on the left parts of the magnons or equivalently on the right parts. This is because the left and right S-matrices are transposed to each other w.r.t.~the left-right inner product.
} \la{LRactio} \vspace{-.2cm} 
\end{figure}
This equivalence can also be phrased by saying that the left and right S-matrices are transposed to each other
\beq
\mathcal{S}_{12}\cdot \chi^{A}_{1}\chi^{B}_{2}\big |\chi^{\dot{A}}_{1} \chi^{\dot{B}}_{2} = \chi^{A}_{1} \chi^{B}_{2}\big| \dot{\mathcal{S}}_{12}\cdot \chi^{\dot{A}}_{1}\chi^{\dot{B}}_{2}\, ,
\eeq
w.r.t.~to the left-right inner product
\beq
\chi^{A}_{1}|\chi^{\dot{A}}_{1} \equiv \mathfrak{h}_{1}^{A\dot{A}}\, , \,\,\, \chi^{A}_{1}\chi^{B}_{2}|\chi^{\dot{B}}_{2}\chi^{\dot{A}}_{1} \equiv \mathfrak{h}_{1}^{A\dot{A}}\mathfrak{h}_{2}^{B\dot{B}}\, , \,\,\, \ldots \, .
\eeq
This fact, together with the unitarity of the S-matrix and the scalar relation~(\ref{Watscal}), guarantee that the form factor~(\ref{2pt}) satisfies the Watson equation.

Consider now the string picture. In this case there is no $Z$ marker, in the conventions of~\cite{AF-string, reviewHuge} that we shall follow. The fundamental representation $\mathcal{V}(p, \zeta)$ is also typically chosen to be unitary and parameterized by two labels only : the momentum $p$ and phase $\zeta$ (i.e.~$|\zeta| = 1$). A generic $n$ magnons state is then given by a tensor product of $n$ bi-modules
\beq\label{z-sequence}
\mathcal{V}(p_{1}, \zeta_{1})\otimes \tilde{\mathcal{V}}(p_{1}, \dot{\zeta}_{1}) \otimes \mathcal{V}(p_{2}, \zeta_{2})\otimes \tilde{\mathcal{V}}(p_{2}, \dot{\zeta}_{2})\otimes \ldots\, ,
\eeq
where $\mathcal{V}$ and $\tilde{\mathcal{V}}$ denote modules for the left and the right $PSU(2|2)$ sharing a common phase $\zeta_i=\dot{\zeta}_i$. The representations are ordered by the condition that $\zeta_{i+1} = e^{ip_{i}}\zeta_{i}$, see~\cite{AF-string, reviewHuge}.

The vanishing of the diagonal central charge~(\ref{P-constraint}) fixes the overall scale of the $\zeta$-sequence in~(\ref{z-sequence}) such that
\beq\label{string-constraint}
\zeta_{1}\dot{\zeta}_1 = \kappa^2 e^{-ip}\, ,
\eeq
with $p=p_1+\ldots + p_n$ the total momentum. (A convenient choice for $\kappa$, which is consistent with $|\zeta_i| = 1$, is then $\kappa = 1$.) In the case of a single magnon $p=p_1$ and the condition~(\ref{string-constraint}) is all we need to define a diagonal $PSU(2|2)$ invariant inner product between left and right modules. The 1-point form factor is of the same form as before, see~(\ref{1-point}), with
\beq\label{Nstring}
N = i\, .
\eeq
For a bigger chain we cannot contract so easily the modules, since the conditions $\zeta_i\dot{\zeta}_i = \kappa^2 e^{-ip_i}$ are not generally met. Remarkably, these conditions are found to be observed after scattering the left (or right) modules. For illustration, for two particles we have
\beq
\mathcal{S} : \mathcal{V}(p_{1}, \zeta_{1})\otimes \mathcal{V}(p_{2}, \zeta_{2}) \rightarrow \mathcal{V}(p_{2}, \zeta'_{2})\otimes \mathcal{V}(p_{1}, \zeta'_{1})
\eeq
where $\zeta'_{2} = \zeta_{1}$ and $\zeta'_{1} = e^{ip_{2}}\zeta'_{2}$, from which one can see that the product of left and right phase now verifies $\zeta'_{i}\dot{\zeta}_{i} = \kappa^2 e^{-ip_i}$. This is saying that we can define an invariant two-point form factor by first scattering the left (or equivalently right) parts and then contracting using the inner product. The two-point form factor in the string frame is thus  the same as in~(\ref{final-app}) except that one should use the string frame S-matrix amplitudes for $A_{12}, \ldots$ as well as the string normalization factor~(\ref{Nstring}). Mind that this makes some differences, since, as well known, the scattering amplitudes in the string frame are not all identical to the ones in the spin chain frame (see also appndix \ref{mapping}). For instance, for scalars,
\beq
A_{12}\big|_{\textrm{string}} = e^{\frac{i}{2}p_{1}-\frac{i}{2}p_2}A_{12}\big|_{\textrm{spin}}\, . \la{scalarExample}
\eeq
For derivatives, however, $D_{12}$ is the same in both frames. The $A/D$ ratio reported in~(\ref{ADratio}) applies in the spin chain frame with normalization $N=1$. Using~(\ref{scalarExample}), this ratio would come with the extra factor $-(x_{1}^{+}x_{2}^{-}/x_{1}^{-}x_{2}^{+})^{\frac{1}{2}}$, in the string frame with $N=i$.

\section{Crossing transformation}\label{cross-app}

The crossing transformation that maps an excitation from one edge of the hexagon to another is most easily understood in the string frame. In this appendix, we derive the crossing transformation rule (\ref{crossingrule}) in that specific frame.
To this end, it is useful to describe magnons in terms of the diagonal $PSU(2|2)_D$ symmetry. A similar analysis was performed in the study of local operators on a Wilson loop in~\cite{DrukkerCorrea}. In that case the right part of the magnon was interpreted as an excitation with reflected rapidity $-u$, whereas in our case the right part will be regarded as an anti-particle. 

Under the full $PSU(2|2)^2$ symmetry, each magnon transforms in a bi-fundamental representation characterized by the quantum numbers,
\beq
\begin{aligned}
&a= \sqrt{g\zeta}\eta,&&b= \sqrt{g\zeta}\frac{i}{\eta}\left( \frac{x^{+}}{x^{-}}-1\right),\\
&c=-\sqrt{\frac{g}{\zeta}}\frac{\eta}{ x^{+}},&&d= -\sqrt{\frac{g}{\zeta}}\frac{x^{+}}{i\eta}\left( \frac{x^{+}}{x^{-}}-1\right).
\end{aligned}
\eeq
where $\eta$ is given by
\beq
\eta=\left( \frac{x^{+}}{x^{-}}\right)^{1/4} \sqrt{i(x^{-}-x^{+})}.
\eeq
When computing hexagon form factors, the parameter $\zeta$ is fixed to be (\ref{string-constraint}), which, for one particle, reads $\zeta = \kappa \sqrt{x^{-}/x^{+}}$.
  
For the left part of the magnon, the diagonal generators act exactly in the same way as the left $PSU(2|2)$ generators:
\beq
\begin{aligned}
\mathcal{Q}^{\alpha}_{\,\,\,\, a}|\phi^{b}\rangle&=a\delta^{b}_{a}|\psi^{\alpha}\rangle, &&\mathcal{Q}^{\alpha}_{\,\,\,\, a}|\psi^{\beta}\rangle=b\epsilon^{\alpha\beta}\epsilon_{ab}|\phi^{b}\rangle,\\
\mathcal{S}^{a}_{\,\,\,\, \alpha}|\phi^{b}\rangle&=c\epsilon^{ab}\epsilon_{\alpha\beta}|\psi^{\beta}\rangle,&&\mathcal{S}^{a}_{\,\,\,\, \alpha}|\psi^{\beta}\rangle=d\delta^{\beta}_{\alpha}|\phi^{a}\rangle.
\end{aligned}
\eeq
Thus, the quantum numbers of the left $PSU(2|2)$ can be naturally identified with the quantum numbers of the diagonal $PSU(2|2)$. On the other hand, for the right part of the magnon, the action of the diagonal generators appears to be non-standard since the roles of $Q$'s and $S$'s are exchanged (\ref{diagonal}):
\beq
\begin{aligned}
\mathcal{Q}^{\alpha}_{\,\,\,\, a}|\phi^{\dot{b}}\rangle&=i\mm c\delta^{\dot{b}}_{a}|\psi^{\alpha}\rangle, &&\mathcal{Q}^{\alpha}_{\,\,\,\, a}|\psi^{\dot{\beta}}\rangle=i\mm d\epsilon^{\alpha\dot{\beta}}\epsilon_{ab}|\phi^{b}\rangle,\\
\mathcal{S}^{a}_{\,\,\,\, \alpha}|\phi^{\dot{b}}\rangle&=\frac{i a}{\mm}\epsilon^{a\dot{b}}\epsilon_{\alpha\beta}|\psi^{\beta}\rangle, &&\mathcal{S}^{a}_{\,\,\,\, \alpha}|\psi^{\dot{\beta}}\rangle=\frac{i b}{\mm}\delta^{\dot{\beta}}_{\alpha}|\phi^{a}\rangle.
\end{aligned}
\eeq
Although non-standard at first glance, one can verify that this is nothing but the standard transformation rule for the excitation with the crossed rapidity $u^{-2\gamma}$ using the property of $\eta$: $\eta^{-2\gamma}=-i\eta/x^{+}$. Therefore, in sum, each magnon transforms in the tensor representation of $PSU(2|2)_D$,
\beq
\mathcal{V}_D(p,\kappa e^{-ip/2})\otimes \mathcal{V}_D(p^{-2\gamma},\kappa e^{-ip^{-2\gamma}/2}).
  \eeq
To be more explicit, each magnon is expressed by a pair of fundamental representations of the diagonal $PSU(2|2)$ symmetry as follows:
\beq
\begin{aligned}\label{diagiden}
\Phi^{a\dot{b}}(u)&\mapsto \phi_D^{a}(u)\phi_D^{\dot{b}}(u^{-2\gamma})\,,\\
\mathcal{D}^{\alpha\dot{\beta}}(u)&\mapsto \psi_D^{\alpha}(u)\psi_D^{\dot{\beta}}(u^{-2\gamma})\,,\\
\Psi^{a\dot{\beta}}(u)&\mapsto \phi_D^{a}(u)\psi_D^{\dot{\beta}}(u^{-2\gamma})\,,\\ \Psi^{\alpha\dot{b}}(u)&\mapsto \psi_D^{\alpha}(u)\phi_D^{\dot{b}}(u^{-2\gamma})\,.
\end{aligned}
\eeq 
 
After the analytic continuation $u\to u^{2\gamma}$, the right hand sides of (\ref{diagiden}) become pairs of excitations with rapidities $u^{2\gamma}$ and $u$. The excitations with $u^{2\gamma}$ can be further replaced by the ones with $u^{-2\gamma}$ using the relation
\beq\label{2gamma-2}
\phi^{a}(u^{2\gamma})=-\phi^{a}(u^{-2\gamma})\,,\quad \psi^{\alpha}(u^{2\gamma})=\psi^{\alpha}(u^{-2\gamma})\,.
\eeq
The relation (\ref{2gamma-2}) can be derived by comparing the actions of $PSU(2|2)$ generators on the two states: In the string frame, their actions are identical up to some minus signs, and the simplest way to eliminate such signs is to identify the states as (\ref{2gamma-2}) \cite{alternativesign}. (This extra minus sign results in a nontrivial monodromy of the S-matrix in the string frame, $S(u^{4\gamma},v)=\Sigma_1 S(u,v)\Sigma_1$ with $\Sigma_1 = {\rm diag}(1,1,-1,-1)$. See e.g.~\cite{AF-string, reviewHuge}.)

After the use of (\ref{2gamma-2}) and reordering, we recover pairs of the excitations with $u$ and $u^{-2\gamma}$. For instance, for scalars and derivatives, we have
\beq\label{reorder}
\begin{aligned}
&\Phi^{a\dot{b}}(u^{2\gamma})\mapsto -\phi_D^{\dot{b}}(u)\phi_D^{a}(u^{-2\gamma})\\
&\mathcal{D}^{\alpha\dot{\beta}}(u^{2\gamma})\mapsto -\psi^{\dot{\beta}}_D(u)\psi^{\alpha}_D(u^{-2\gamma})
\end{aligned}
\eeq
(Note that the overall minus sign for the derivatives comes from reordering of fermions.) Importantly, the right hand sides of (\ref{reorder}) are identical to those of (\ref{diagiden}) except for the exchange of the left and the right $PSU(2|2)$ indices and the extra minus sign. We thus conclude that the excitation transforms under the analytic continuation $u\to u^{2\gamma}$ as  
\beq
\chi^{A\dot{B}}\overset{2\gamma}{\to}-\chi^{B\dot{A}}\,.
\eeq

\section{Useful Mirror and Crossing Formulae} \la{crossingFormulae}

\begin{figure}[t]
\begin{center}
\includegraphics[scale=0.31]{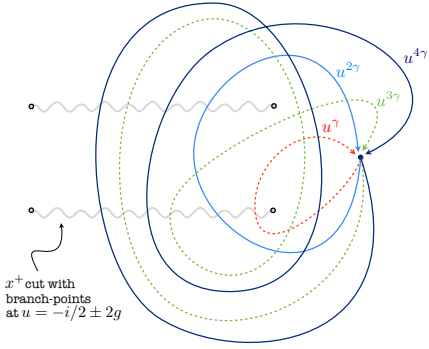}
\end{center}
\vspace{-2cm}
\caption{\normalsize Mirror and crossing transformations correspond to transporting the rapidity along paths that cross the $x^+$ and $x^-$ Zhukowsky cuts. A crossing transformation $u\to u^{2\gamma}$ transforms a particle into its anti-particle and a full monodromy $u \to u^{4\gamma}$ brings it back to its original form.
} \la{monodromies} \vspace{-.2cm} 
\end{figure}

The mirror transformations are monodromies in the $u$ plane which take $u$ back to the same value but on a different sheet, see figure \ref{monodromies}. Under such monodromies, the Zhukowsky variables transform as:
\beqa
&&u\to u^\gamma \, \,\, :\qquad  x^- \to x^- \,\,\,\,\,\,\, , \qquad  x^+ \to 1/x^+ \,,\\
&&u\to u^{2\gamma} \, :\qquad  x^- \to 1/x^- \, , \qquad  x^+ \to 1/x^+ \,,\\
&&u\to u^{3\gamma} \, :\qquad  x^- \to x^- \,\,\,\,\,\,\, , \qquad  x^+ \to 1/x^+ \,, \\ 
&&u\to u^{4\gamma} \, :\qquad  x^- \to x^- \,\,\,\,\,\,\, , \qquad  x^+ \to x^+ \,.
\eeqa
Two mirror transformations is what one denotes as a crossing transformation which transforms a particle into an anti-particle. After four mirror (i.e.~two crossing) transformations the particle comes back to itself. For simple functions of $u$, with a simple dependence on the Zhukowsky variables -- like the entries of the matrix part or for transfer matrices -- four mirror transformations have no effect. For such quantities $u^{3\gamma}$ is undistinguishable from the inverse mirror transformation $u^{-\gamma}$. 
%
On the other hand, for more involved quantities -- such as the dressing phase or the dynamical part of our hexagon for factors which contains the dressing phase -- one should be more careful.

Indeed, the dressing phase $\sigma$ has non-trivial monodromies even under $u\to u^{4\gamma}$. If we are interested in moving physical particles from the top of the hexagon to one of the other two physical edges all we need is the crossing transformation
\beq
\sigma(u^{2\gamma}, v)\sigma(u, v) = \frac{(1-1/x^+y^{+})(1-x^-/y^{+})}{(1-x^-/y^{-})(1-1/x^+y^{-})} 
\eeq
combined with unitarity $\sigma(u,v)=1/\sigma(v,u)$. With these, we can readily obtain 
\beqa
\!\!\!\!\!\!\!\!\!\!\!\!\! h_{DD}(u\,\,\,\,\,\,,v)&=&\frac{x^{-}-y^{-}}{x^{-}-y^{+}} \frac{1-1/x^{-}y^{+}}{1-1/x^{+}y^{+}} \frac{1}{\sigma(u,v)} \,, \la{l1} \\
\!\!\!\!\!\!\!\!\!\! \!\!\! h_{DD}(u^{2\gamma},v) \la{l2}
&=&\frac{1-1/x^+y^{-}}{1-1/x^{-}y^+} \sigma(u,v) \la{l3}  \,, \\
\!\!\!\!\!\!\!\!\!\! \!\!\! h_{DD}(u^{4\gamma},v)&=&\frac{y^{-}-x^{+}}{y^{-}-x^{-}} \frac{1-1/y^{+}x^{+}}{1-1/y^{-}x^{+}} \frac{1}{\sigma(u,v)}\,,\\
\!\!\!\!\!\!\!\!\!\! \!\!\! h_{DD}(u^{6\gamma},v)&=& \frac{x^- - y^+}{x^- - y^+} \frac{1-1/x^+ y^-}{1-1/x^+ y^+} h_{DD}(v,u) \la{l4}
\eeqa
from where we obtain that $h_{DD}(u^{4\gamma}, v) = 1/h_{DD}(v, u)$. Finally, it is also useful to notice that $h_{DD}(u^{2\gamma}, v^{2\gamma}) = h_{DD}(u, v)$, which expresses the invariance of the two-point function upon crossing of the full state to another edge.
 
  \begin{figure}[t]
\begin{center}
\includegraphics[scale=0.31]{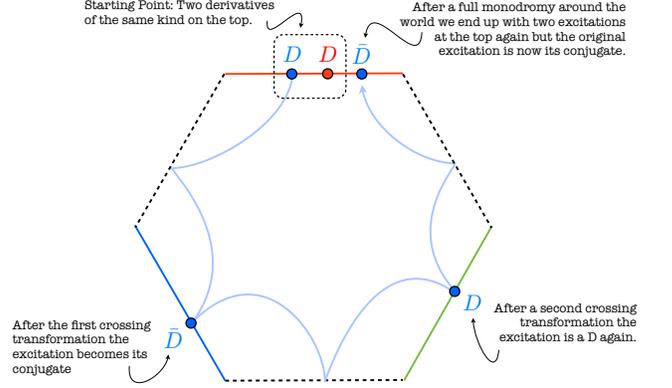}
\end{center}
\vspace{-1.5cm}
\caption{\normalsize A particle goes around the world through a sequence of 6 mirror rotations. Every two steps its flavour gets swapped $\chi^{A\dot{B}} \rightarrow - \chi^{B\dot{A}}$. For a longitudinal derivative $D  = \mathcal{D}^{1\dot{2}}$, it means that $D\rightarrow -\bar{D}$ upon crossing, as for a round trip around the hexagon.
} \la{monodromies2} \vspace{-.2cm} 
\end{figure}

 As expressed in~(\ref{l1}--\ref{l4}), the weak coupling expansion of each of these quantities is straightforward. To leading order it suffices to use $x^\pm \simeq (u\pm i/2)/g$ and $\sigma \simeq 1$.

Let us end this appendix with a few small comments on some nice features of these formulae, see also figure \ref{monodromies2}. First we recall that after a crossing transformation a particle transforms into its conjugate. As such, the second line (\ref{l2}) describes a direct transition where a $D$ on the top edge becomes, up to a sign, a $\bar D$ on the left bottom edge of the hexagon. 
Such process should be invisible to leading order since a propagator can never connect such derivatives. In perfect agreement with this expectation, we observe neatly that the second line trivializes at weak coupling, 
$$h_{D|\bar D}(v|u) \equiv -h_{DD}(u^{2\gamma},v) = -1 + \mathcal{O}(g^2) \,.$$ 
On the other hand, as we perform a second crossing and move to the third line (\ref{l3}), the derivative is again a $D$. A process with a derivative in an edge propagating to a bottom edge ought to show up at leading order in perturbation theory and indeed, here we have a non-trivial result
$$h_{D| D}(v|u) \equiv h_{DD}(u^{4\gamma},v) = \frac{u-v+i}{u-v} + \mathcal{O}(g^2) \,.$$ 
already at leading order. As clearly seen in (\ref{l2}) and (\ref{l3}), a further important difference between the two direct transitions just discussed is that the latter has a pole for $u = v$ while the former does not. The pole in the direct transition $h_{D|D}$ yields the measure for the $D$ excitations. As we cross three times, as in the last line (\ref{l4}), we are back to the original edge after an odd number of crossing. As such, the derivative arrives back at this edge to the right of the excitation $v$ which stayed there and -- most importantly -- as a conjugate excitation $\bar D$. In other words, we expect $h_{DD}(u^{6\gamma},v) = h_{D\bar{D}}(v,u)$ which is not the same as $h_{D{D}}(v,u)$. Nonetheless, the ratio between the two -- made explicit in (\ref{l4}) -- is fixed by symmetry and is thus something we can now non-trivially verify. More precisely, from the middle line of (\ref{all2pt}) we expect 
\beq
h_{D \bar D}(u,v)/h_{D D}(u,v)= \frac{D_{12}-E_{12}}{2D_{12}}
\eeq
which perfectly matches the pre-factor in (\ref{l4}). Note that all these simple considerations are already probing the finite coupling structure of our proposals even if we end up comparing such transitions to weak coupling data. The reason is that all these direct transitions and creation amplitudes are all non-trivially related by crossing transformations which, in turn, do not commute with perturbation theory. 


%
%
%



All of the above concerned fundamental excitations. For bound-states a mirror transformation is a bit more involved:  $u\rightarrow u^{\pm \gamma}$ is directly obtained by crossing $x^{[\pm a]}\rightarrow 1/x^{[\pm a]}$, while keeping fixed the remaining Zhukowsky variables. 

\section{Mapping spin-chain and string frames}\label{mapping}

In this appendix we discuss the mapping between spin-chain and string frame excitations, in the particular case of bosonic excitations. The string frame is especially well suited for crossing excitations, since in this case the recipe boils down to swapping the Zhukowsky variables and slightly rearranging the indices as explained in appendix~\ref{cross-app}. On the other hand, the spin-chain frame greatly simplifies the comparison with perturbative data at weak coupling. It is then useful to be able to navigate between the two.

To make the dictionary as leaner as possible, we shall specialize to a particular set of spin chain parameters. Namely, we shall fix $\kappa = i\alpha = 1$ as well as a normalization for which $N=i$. In this case the mapping takes the form
\beq\la{dicbos}
\mathcal{D}_{\textrm{string}} = \mathcal{D}_{\textrm{spin}}\, , \qquad \Phi_{\textrm{string}} = \sqrt{Z}\Phi_{\textrm{spin}}\sqrt{Z}\, ,
\eeq
and, given the rule for moving the $Z$ marker around, i.e.~$\chi Z = e^{ip}Z\chi$, it is simple algebra to derive the general relation between (bosonic) multi-particle states in the two frames. The most interesting case corresponds to having $n$-scalar fields, for which we find
\beq\label{map}
\big|\Phi_{1}\ldots \Phi_{n}\big>_{\textrm{string}} =\prod_{i}\frac{e^{ip_i/2}\zeta_{i}}{\zeta} \big|Z^{n}\Phi_{1}\ldots \Phi_{n}\big>_{\textrm{spin}}
\eeq
where $\zeta_i$ are the (nonlocal) parameters of the chain of magnons, see~(\ref{z-sequence}), and where $\zeta = e^{-\frac{i p}{2}}$ solves~(\ref{string-constraint}) with $\kappa=1$.

The main motivation for~(\ref{dicbos}) is that it guarantees proper mapping between spin chain and string frame S-matrix. As a simple cross-check, consider scalar scattering, which in both frames can be written
\beq
\mathcal{S}_{12}\big|\Phi^{1\dot{1}}_{1}\Phi_{2}^{1\dot{1}}\big> = S_{12}^0 A_{12}\dot{A}_{12}\big|\Phi^{1\dot{1}}_{2}\Phi_{1}^{1\dot{1}}\big>\, ,
\eeq
with the same scalar factor $S_{12}^0$. (As in \cite{Beisert,Beisert:2006qh}, we ignore $Z$ markers at the left end of the chain when we compute the S-matrix.)
Using the relation~(\ref{map}), we get
\beq
A_{12}\dot{A}_{12}\big|_{\textrm{spin}} = \frac{\zeta'_{1}\zeta'_{2}}{\zeta_{1}\zeta_{2}} A_{12}\dot{A}_{12}\big|_{\textrm{string}}\, ,
\eeq
where $\zeta'_i$ denotes values after scattering. Using $\zeta_{2} = e^{ip_{1}}\zeta_{1} = e^{ip_{1}}\zeta$ and $\zeta'_{1} = e^{ip_{2}}\zeta'_{2} = e^{ip_{2}}\zeta$, we then recover the well-known relation between string and spin chain frame for the $SU(2)$ S-matrix, see (\ref{scalarExample}).

To derive the relation between hexagon form factors in two frames, we need to use the rule (\ref{z-rule}) for pulling out $Z$ markers, in addition to the dictionary (\ref{dicbos}). For states with scalar excitations, 
\beq
\begin{aligned}\nonumber
&|\psi_1\rangle =|\Phi(p_1)\cdots \Phi(p_l)\rangle,\\
&|\psi_2\rangle =|\Phi(q_1)\cdots \Phi(q_m)\rangle,\\
& |\psi_3\rangle =|\Phi(r_1)\cdots \Phi(r_n)\rangle ,
\end{aligned}
\eeq
the mapping (\ref{dicbos}) leads to
\beq
\begin{aligned}\nn
\langle \mathfrak{h}|\psi_1\rangle|\psi_2\rangle|\psi_3\rangle_{\rm string}&=
F_pF_qF_r\langle \mathfrak{h}|Z^{l}\psi_1\rangle|Z^{m}\psi_2\rangle|Z^{n}\psi_3\rangle_{\rm spin},
\end{aligned}
\eeq
where the factors $F_{p,q,r}$ are given by
\beq\nn
F_{p}=\prod_{k}\frac{e^{ip_k/2}\zeta_k}{\zeta} ,\,\,F_{q}=\prod_{k}\frac{e^{iq_k/2}\zeta_k}{\zeta} ,\,\,F_{r}=\prod_{k}\frac{e^{ir_k/2}\zeta_k}{\zeta} .
\eeq
One can further simplify the expression by using the rule~(\ref{z-rule}) for pulling out $Z$'s. However, since (\ref{z-rule}) is applicable only when all excitations live on a single edge, we need to perform the crossing transformations and gather excitations in one of three kets. This can be achieved by first dressing the spin chain excitations into stringy ones, like e.g.~$\sqrt{Z}\Phi\sqrt{Z}$, using dictionary~(\ref{dicbos}) and then applying the crossing rule in the string frame (\ref{crossingrule}). After doing so, we can apply the rule (\ref{z-rule}) and remove the $Z$ markers from the left edge of that ket. By repeating the same procedures for the remaining two kets, we can get rid of all extra $Z$ factors. Although such manipulation leads to complicated expressions at the intermediate stages, it yields the following simple formula at the end of the day:
\beq
\begin{aligned}\la{hexstsp}
\langle \mathfrak{h}|\psi_1\rangle|\psi_2\rangle|\psi_3\rangle_{\rm string}&=e^{-\frac{i}{2}\left[P(l+n-m)+Q(m+l-n)+R(n+m-l)\right]}\\
&\times F_pF_qF_r\langle \mathfrak{h}|\psi_1\rangle|\psi_2\rangle|\psi_3\rangle_{\rm spin},
\end{aligned}
\eeq
Here $P$, $Q$ and $R$ are the total momenta for each chain.

As an application, let us derive a simple crossing rule in the spin-chain frame for the states with a single scalar excitation. 
In the string frame, the crossing rule takes the simple form as explained in appendix \ref{cross-app}:
\beq
\big<\mathfrak{h}\big|\Phi_{3}\big>\big|\Phi_{2}\big>\big|\Phi_{1}\big>_{\textrm{string}} = -\big<\mathfrak{h}\big|\Phi^{4\gamma}_{1}\tilde{\Phi}^{2\gamma}_{2}\Phi_{3}\big>_{\textrm{string}}\, ,
\eeq
Using the mapping between the hexagon form factors (\ref{hexstsp}), we can translate this to the crossing rule in the spin-chain frame,
\beq\label{spcross}
\big<\mathfrak{h}\big|\Phi_{3}\big>\big|\Phi_{2}\big>\big|\Phi_{1}\big>_{\textrm{spin}} = -e^{ip_1-ip_3} \big<\mathfrak{h}\big|\Phi^{4\gamma}_{1}\tilde{\Phi}^{2\gamma}_{2}\Phi_{3}\big>_{\textrm{spin}}\, .
\eeq
As this example shows, the crossing rule in the spin-chain frame often entails momentum factors, which must be taken into account when comparing with the weak-coupling results (for an explicit example, see appendix~\ref{Further}).

Let us conclude by explaining, on a simple but generic enough example, why the additional factors in the string frame are irrelevant at the level of the three-point function. As well known from the BA equations story, the main point is that the string frame makes use of a different notion of length, the BMN length $J = L-M$ where $M$ is the number of scalars. Said differently, the length in the string frame is the R-charge. 
Take the all loop asymptotic result (\ref{calAformula}) for one non-BPS and two protected operators in the $SU(2)$ sector. 
We have $L^\text{spin} =L^\text{string}+M$ and therefore $\ell_{12}^\text{spin}=\ell_{12}^\text{string}+M/2$ from the definition of the partial length. Also, inherited by the relation above (\ref{hexstsp}),  
\beq
h_{YY}^\text{spin}(u,v)= e^{\frac{i}{2} p(v)-\frac{i}{2}p(u) } h_{YY}^\text{string}(u,v) \,.
\eeq
Using these relations we shall now show that the (non-trivial part of the) summand in (\ref{calAformula}), i.e.
\beq
\prod_{j\in \bar{\alpha}}  e^{ip_{j}\ell^\text{spin}} \prod_{i\in \alpha, j\in \bar{\alpha}} f^\text{spin}(u_{i}, u_{j}) 
\eeq
is frame independent. (Recall that $f=1/h$.) In what follows we use configurations with strict zero total momentum, i.e. no-winding. This avoids dealing with several ambiguities concerning products of square-roots versus square root of products and similar issues. Indeed, it is well known that the string frame is under better control in the absence of winding,  see e.g. discussion around (2.14) in the review \cite{reviewHuge}. Using the relations above we can re-write the summand as
\beqa
&&\!\prod_{j\in \bar{\alpha}}  \( e^{ip_{j}\ell^\text{string}} e^{ip_{j}M/2} \) \!\! \prod_{i\in \alpha, j\in \bar{\alpha}}\!\!\! \( e^{\frac{i}{2} p(u_i)-\frac{i}{2} p(u_j)} f^\text{string}(u_{i}, u_{j})  \) \nn \\
&&\!=\!\Big( \prod_{j\in \bar{\alpha}}  e^{ip_{j}\ell^\text{string}}\!\!\! \prod_{i\in \alpha, j\in \bar{\alpha}} \!\!\! f ^\text{string}(u_{i}, u_{j}) \Big)  \!\times \!e^{\frac{i}{2} P_{\bar{\alpha}} M+\frac{i}{2}  P_{\alpha} |\bar \alpha|-\frac{i}{2}  P_{\bar\alpha} | \alpha|} \nn
\eeqa
where $P_{\alpha}$ is the total momentum of the partition $\alpha$, $P_{\alpha} = \sum_{i \in \alpha} p_i$ while $P_{\bar\alpha}$ is the total momentum of the complement partition $\bar \alpha$. Using $M=|\alpha|+|\bar{\alpha}|$ and the zero momentum condition $P=P_\alpha+P_{\bar\alpha}=0$ we can immediately simplify the last exponential to $1$. In sum: while intermediate quantities are scheme dependent, for physical quantities such as $\mathcal{A}$ we are free to choose whichever frame we want. 

\section{Partitions from a form factor expansion\label{appA}}

\begin{figure}[t]
\begin{center}
\includegraphics[scale=0.31]{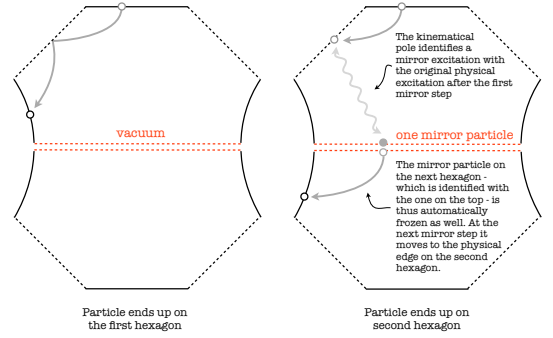}
\end{center}
\vspace{-2.2cm}
\caption{\normalsize 
As we cross an excitation from an edge to another it can interact with the virtual particles which glue together the two hexagons. When it does, it can either stay on the same hexagon or, effectively, teleport to the (same physical edge but on the) neighbouring hexagon. This provides one with an alternative explanation for the origin of the sum over partition of physical excitations. 
} \la{partitions} \vspace{-.2cm} 
\end{figure}

In this appendix we discuss the crossing transformation of an excitation living on a pair of joined hexagons. The motivation for carrying out this study is that it sheds light on the origin of the sum over partitions appearing in figure~\ref{hexagon}.

In the set up we consider we have two hexagons glued together into an octagon of sort through a single mirror channel, as depicted in figure~\ref{partitions} (where the sum over mirror states there is truncated to the vacuum and single particle contributions). 
We will show that the sum over partitions is the expected answer for describing such an octagon dressed with excitations on its sides.

To begin with, we first imagine putting a single excitation on one of the opposing edges, say a magnon $D$ with rapidity $v$. In this case we can immediately write what the answer, i.e.~the corresponding octagon form factor, should be. It starts with the asymptotic contribution
\beq\label{asym}
\textrm{asymptotic octagon} = h^{(1)}_{D}(v)\times h^{(2)}_{\textrm{vac}} = h^{(1)}_{D}(v)\, ,
\eeq
which is the product of two form factors, one for the hexagon hosting the excitation $v$ and the other for the empty hexagon (we set the vacuum amplitude to $1$ in that case). When we bring the two hexagons closer to each other, by reducing the distance $\ell_{23}$ between them, we need to include the tower of finite size corrections, resulting in the infinite series
\beq\label{full}
\begin{aligned}
&\textrm{full octagon} = h^{(1)}_{D}(v) \\
&+ \sum_{X}\int \frac{du}{2\pi} \frac{\mu_{X}(u^{\gamma})}{(x^{+}x^{-})^{\ell_{23}}}h^{(1)}_{XD}(u^{3\gamma}, v)h^{(2)}_{\bar{X}}(u^{-3\gamma}) \\
&+ \textrm{heavier states}\, ,
\end{aligned}
\eeq
where $X$ is a sum over the mirror magnon flavours, $\mu_{X}(u^{\gamma})$ the associated integration measure, and where the form factors describe the creation of an $X\bar{X}$ pair on the two sides of the mirror channel. See similar discussion below~(\ref{calAwrapping}) in section~\ref{FSsection} and figure~\ref{partitions}.

As complicated as the above series can be, it does not involve any sum over partitions. The reason is that we are exciting an edge that stands too far away from the place where we glued our hexagons. On the contrary, the sum over the partitions is the answer when we excite an edge on the side, i.e.~one which can feel the cut mirror channel. To reveal this effect, we can rotate our probe excitation to one of these edges, by performing the crossing transformation $v\rightarrow v^{2\gamma}$. When applied to the asymptotic part~(\ref{asym}) it amounts simply to replacing $v$ by $v^{2\gamma}$.
However, this expression is not the full asymptotic part of the result. Indeed, we also need to pay attention to what is happening under the integral sign in~(\ref{full}). Proceeding more carefully, we first rotate our particle to the adjacent mirror channel $v\rightarrow v^{\gamma}$ and find that the integrand in~(\ref{full}) contains the factor
\beq
h^{(1)}_{DX}(u^{3\gamma}, v^{\gamma})\, .
\eeq
The important fact about it is that it has a pole whenever $u\sim v$ and $X = \bar{D}$. This one is nothing else that the mirror version of the pole encountered earlier in the direct transition, and its residue is directly related to the mirror measure $\mu_{X}(u^{\gamma})$ in~(\ref{full}). As we move the particle along the $\gamma$ path, the pole is approaching the contour of integration from below. If moreover we push this pole further up in the complex $u$ plane, which is what we intend to do for the crossing under study, we should deform the contour of integration and pick up the residue, which reads
\beq
\frac{1}{(y^{+}y^{-})^{\ell_{23}}}\times h^{(2)}_{D}(v^{-3\gamma})\, .
\eeq
What is special about it is that it is \textit{not} always suppressed at large distance $\ell_{23}$. This is the case at the end of the crossing transformation, which is reached after performing $v\rightarrow v^{\gamma}$ one more time and which turns $y^{-}\rightarrow 1/y^{-}$ while leaving $y^{+}$ unchanged. This residue should thus be added to the asymptotic answer, leading to
\beq
h^{(1)}_{D}(v^{2\gamma}) + h^{(2)}_{D}(v^{-2\gamma}) e^{-ip\ell_{23}} + \dots\, ,
\eeq
which is the sum over partitions we were looking for. By this mechanism an excitation which was originally on one hexagon has been teleported to the following one, on the other side of the cut channel. When doing this we do not expect to be able to kill more than 1 integral at a time. Therefore the dots above should be all mirror processes and thus subleading at large distance.

The moral is that putting excitation on an edge intersecting a cut channel requires decomposing the wave function of this excitation into two parts, describing the before and the after of this cut. Iterating this procedure starting with more particles and crossing them one after the other leads to the sum over partitions. (Note that this sum is not obtained for free and actually puts constraints on the structure of the multi-particle integrands, which should have appropriate singularities. We expect these ones to be in place thanks to the decoupling property of our multi-particle ansatz.)

\section{Transfer matrices} \la{tm}

Transfer matrices in symmetric and antisymmetric representations were computed in~\cite{Beisert:2006qh}. For our purposes, only the latter ones are relevant. They take the form
\beq
\begin{aligned}
T_{a}(u) &= (-1)^a\sum_{n=-1}^{1}(3n^2-2)\prod_{m=0}^{n}\frac{R^{(+)}(u^{[2m-a]})}{R^{(-)}(u^{[2m-a]})}\times \\
&\,\,\sum_{j=\frac{2-a}{2}}^{\frac{a-2n}{2}}\prod_{k=j+n}^{\frac{a-2}{2}}\frac{R^{(+)}(u^{[2n-2k]})B^{(+)}(u^{[-2k]})}{R^{(-)}(u^{[2n-2k]})B^{(-)}(u^{[-2k]})}\, ,
\end{aligned}
\eeq
where
\beq
R^{(\pm)}(u) = \prod_{j}(x(u)-x^{\mp}_{j})\, , \,\,\, B^{(\pm)}(u) = \prod_{j}(\frac{1}{x(u)}-x^{\mp}_{j})\, .
\eeq
Its mirror version $u\rightarrow u^{\pm \gamma}$ is directly obtained by crossing $x^{[\pm a]}\rightarrow 1/x^{[\pm a]}$, while keeping fixed the remaining Zhukowsky variables. For $a=1$ it takes the simple form
\beq
T(u) = 1+\prod_{j}\frac{x^{-}-x^{-}_{j}}{x^{-}-x^{+}_{j}} \frac{1-1/x^{+}x^{-}_{j}}{1-1/x^{+}x^{+}_{j}}-2\prod_{j}\frac{x^{-}-x^{-}_{j}}{x^{-}-x^{+}_{j}}\, .
\eeq
In the semiclassical strong coupling limit, we can introduce a density of roots and write 
\beq
\begin{aligned}
T(u) =&\,\, 1+\exp{i\int \frac{dv}{g}\rho(y)\frac{xy}{(x-y)(xy-1)}} \\
& -2\exp{i\int \frac{dv}{g} \rho(y)\frac{y^2}{(x-y)(y^2-1)}}\, .
\end{aligned}
\eeq
It is equivalent to
\beq
T(u) = e^{i\hat p(x)} (e^{-i\hat p(x)}+e^{i\hat p(1/x)}-2e^{-i\tilde p(x)})\, , \la{Tp}
\eeq
where $\hat p(x)$ is the Anti-de-Sitter semiclassical momentum
\beq
\hat p(x) = \frac{x L}{2g(x^2-1)} + \int \frac{dv}{g}\rho(y)\frac{xy(xy-1)}{(x^2-1)(x-y)(y^2-1)}\, ,
\eeq
and $\tilde p(x)$ the sphere counterpart
\beq
\tilde p(x) = \frac{x L}{2g(x^2-1)} -\frac{2\pi m}{x^2-1}\, ,
\eeq
with $m$ the winding number (i.e.~the value of the total spin-chain momentum). In the text we were considering $m=0$ for simplicity.

\section{More on $\mathcal{A}$} \la{complementCalA}


A priori, it is not clear whether the expression (\ref{calAformula}) for $\mathcal{A}$ is in line with our hexagon picture. After all, we would expect to have interactions between roots within the same partition (corresponding to excitations in the same hexagon) rather than interactions between different partitions as in the product of $f(u_{i}, u_{j}) $ in (\ref{calAformula}). The point is that the product of $h(u_{i}, u_{j})$ outside the sum precisely brings it to the desired form with the expected pre-factors. As illustration, it is instructive to work out explicitly the simple case of two excitations. We get
\beq
\!\! \mathcal{A} \!=\! h_{12} \!\[ 1\! -\! f_{12} e^{i p_2 \ell}\! -\! f_{21} e^{i p_1 \ell} \!+\! e^{i p_1 \ell +i p_2 \ell} \] \, ,\nn
\eeq
where $h_{12} = h(u_{1}, u_{2})$ and similarly for $f$. Using $h_{12}=1/f_{12}$ and $f_{12}/f_{21}=S_{21}$, this simplifies to  
\beq
\mathcal{A} \!=\! h_{12}  -  e^{i p_2 \ell}\! -\! S_{12} e^{i p_1 \ell} \!+\! h_{12}  e^{i p_1 \ell +i p_2 \ell}\, .  \la{calAformula2Magnons}
\eeq
Each term has now a transparent interpretation. The first one corresponds to having both excitations on the first hexagon. The second term corresponds to the second magnon having moved to the second hexagon (hence the exponential) while the first magnon remained in the first hexagon. This second term should therefore be given by a product of two hexagon one point functions which we normalized to one. The third term is similar with the first and second magnon swapped; the additional S-matrix factor precisely takes the extra phase shift acquired by particle $1$ (originally to the left) as it scatters through particle $2$. Finally, the last term is given by both excitations ending up on the second hexagon. 

In the end, we could have massaged (\ref{calAformula}) into a form closer to (\ref{calAformula2Magnons}) whose physical interpretation would be clearer. We chose to kept it as is to make contact with previous expressions in the literature.

Further on the comparison with previous results, let us conclude this appendix with a few comments concerning the measure $\mu(u)$ in (\ref{rank1}). In \cite{tailoring1,tailoring4,tailoringNC} --  where the structure constants (\ref{rank1}) were first computed up to one loop level --  the pre-factor multiplying the sum over partitions took a seemingly different form. It is however very simple to realize that it is precisely equivalent to what we propose to this loop order. Take for example equations (12), (13) and (14) in \cite{tailoringNC}. We immediately observe a perfect agreement with (\ref{rank1}) except that we would read off the pre-factor 
\beq
\texttt{pre-factor}= \frac{1-\frac{1}{2} \sum\limits_{j=1}^S \epsilon(u_j)}{\sqrt{1/\prod\limits_{j=1}^S \frac{\partial p(u_j)}{u_j}} \,\prod\limits_{j=1}^S(e^{-ip_j}-1) } \la{exampleNC}
\eeq
rather than the product of measures as in (\ref{rank1}). However, to this loop order, $\epsilon(u)=O(g^2)$ and the numerator in (\ref{exampleNC}) can therefore also be cast as a product $\prod_j \(1-\frac{1}{2}  \epsilon(u_j)\)$ and combined with the other products. In the end we get
\beq
| \texttt{pre-factor} |=  \prod_{j=1}^S \sqrt{1-\frac{g^2}{(u_j^2+\frac{1}{4})^2} +\mathcal{O}(g^4)}
\eeq 
which neatly agrees with the perturbative expansion of the measure
\beq
 \mu(u)= \frac{i}{\underset{v=u}{\operatorname{{\rm residue}}} \,f(u,v)}=1-\frac{
   g^2}{\left(u^2+\frac{1}{4}\right)^2}-\frac{g^4\! \left(8 u^2-1\right)}{\left(u^2+\frac{1}{4}\right)^4} +\dots  \nn \,.
\eeq
A similar analysis could be done for the $SU(2)$ structure constants of \cite{tailoring1,tailoring4}. Again, we find perfect agreement with all previously known results.

\section{Harmonic sums representation}

Perturbative corrections to the ratio $C_{123}^{\bullet \circ \circ}/C_{123}^{\bullet \circ \circ}|_{\textrm{tree}}$ can be written, at low twist, in terms of nested harmonic sums, as can be checked explicitly up to three loops by inspecting the known expressions~\cite{Eden:2012rr}. We give here in such a basis the representation of the finite-size correction $\delta\mathcal{A}_{23}$ for a size-one bridge, i.e.~$\ell_{23} = 1$. We found (by setting up an ansatz and fixing up its coefficients by numerical comparison with Eqs.~(\ref{upshot}),~(\ref{I}),~(\ref{J})) that
\beq
\begin{aligned}
&\frac{\delta\mathcal{A}_{23}}{4g^4\mathcal{A}} = S_{-2}^2-2S_{-3}S_{1}-2S_{-2}S_{1}^2-2S_{1}S_{3}+2S_{-3, 1}\\
&-S_{4}+4S_{-2, 1}S_{1}+2S_{-2, 2}+2S_{3,1}-4S_{-2, 1, 1} +O(g^2)\, ,
\end{aligned}
\eeq
with the usual definition for the nested sums, see e.g.~\cite{Alday:2013cwa,Eden:2012rr}, and with the spin $S$ entering as their argument. In this form, we can easily read out the few first terms in the large spin expansion,
\beq
\frac{\delta\mathcal{A}_{23}}{4g^4\mathcal{A}} = \frac{\pi^2}{6} S_{1}^2 -3\zeta_{3}S_{1} +\frac{7\pi^4}{360} + O(1/S)\, ,
\eeq
with $S_{1} = \log{S}+\gamma_{E} + O(1/S)$. This shows that finite-size corrections to structure constants are not negligible at large spin, in contrast with what happens for scaling dimensions.

\section{Off-shell scalar product from hexagons }\label{Offshell}
\begin{figure}[t]
\begin{center}
\includegraphics[scale=0.3]{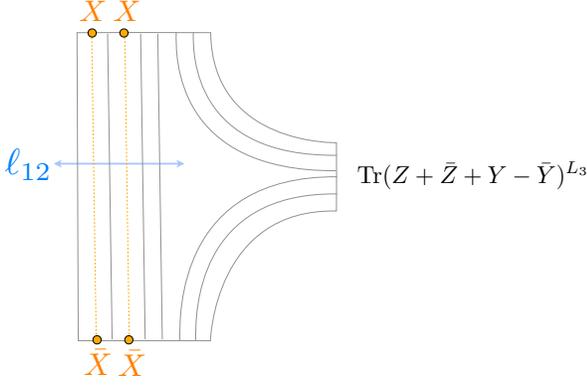}
\end{center}
\vspace{-1.5cm}
\caption{\normalsize 
The tree level structure constant is given by the off-shell scalar product of the XXX spin chain with length $\ell_{12}$. Note that the configuration studied here is different from the ones studied in \cite{tailoring1, tailoring3, Ivan2, ivan, tailoring4, Kazama:2014sxa}: Although non-BPS operators contain $X$ and $\bar{X}$ excitations as in \cite{tailoring1}, the third operator, which acts as a reservoir, contains $Y$ and $\bar{Y}$ instead of $X$ and $\bar{X}$. Therefore the result cannot be simplified further using the existing techniques \cite{Ivan2,Kazama:2014sxa}.
} \la{XXex} \vspace{-.2cm} 
\end{figure}

In this appendix we illustrate how the hexagon vertex approach makes contact with more direct weak coupling computations of three-point correlators in terms of spin chain scalar products \cite{roiban,tailoring1}. 

We consider here a three-point function of one BPS operator and two non-BPS $SU(2)$ operators; one with $X$ excitations with rapidities $u_j$ and the other with $\bar{X}$ excitations with corresponding rapidities $v_j$, see figure~\ref{XXex}. The scalars talk to each other through the bridge of size $\ell_{12}$ connecting the two non-BPS operators such that this correlator can be expressed in terms of the off-shell scalar product of the XXX spin chain with length $\ell_{12}$.  The off-shell scalar products were first computed by Korepin \cite{korepin} as a double sum over bi-partite partitions, see also \cite{Slavnov}and the equation (114) in~\cite{tailoring1}.

From the hexagon approach point of view, a double sum over partitions arises from the get-go from cutting the pair of pants into two hexagons and summing over which excitations of either operator end up on which hexagon, 
%
%
 \beq
 \sum_{\substack{\alpha\cup \bar{\alpha}=\{u\}\\\beta\cup\bar{\beta}=\{v\}}} w_{\ell_{13}}(\alpha,\bar \alpha) w_{\ell_{12}}(\beta,\bar \beta) 
 \, h(\alpha|\beta)\, h(\bar{\beta}|\bar{\alpha})\label{XXbar3pt}
 \eeq
 with the splitting pre-factors (here $a_\ell(u) \equiv e^{i p(u) \ell}$)
%
%
 \beqa
w_{\ell_{13}}(\alpha,\bar \alpha) &=& \prod_{ u_j \in \bar \alpha} (a_{\ell_{13}}(u_j)    \prod_{u_i \in \alpha \text{ with }  i > j}  S(u_j,u_i) )  \nn \\
&=& \prod_{ u_j \in \bar \alpha} (a_{\ell_{12}}^{-1}(u_j)  
\prod_{u_i \in \alpha \text{ with }  i < j}  S^{-1}(u_j,u_i) ) \,, \nn \\
w_{\ell_{12}}(\beta,\bar \beta) &=& \prod_{ v_j \in \bar \beta} (a_{\ell_{12}}(v_j)  \prod_{v_i \in \beta \text{ with }  i > j}  S(v_j,v_i) ) \,. \nn
\eeqa
Note that for $w_{\ell_{13}}$ we used Bethe equations to eliminate $\ell_{13}$ for $\ell_{12}$. This alternative representation is the convenient one if we want to recognize (\ref{XXbar3pt}) as the off-shell scalar product alluded to above (which only knows about~$\ell_{12}$). 

\begin{figure}[t]
\begin{center}
\includegraphics[scale=0.3]{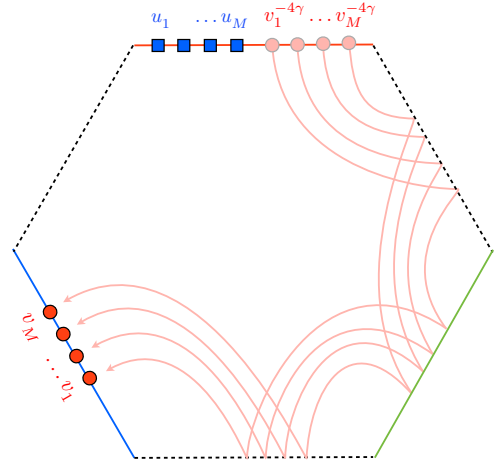}
\end{center}
\caption{\normalsize To obtain a correlator with $M$ scalars $X$ at the top and $M$ conjugate scalars $\bar X$ at the bottom we start with a creation amplitude with $M$ $X$'s and $M$ $\bar X$'s at the top and rotate the latter by means of crossing transformations. Here we choose to perform two crossing transformations on the right (rather than just one on the left) to simplify the math. Indeed, under $v \to v^{-4\gamma}$, the matrix part is invariant rendering contact with the off-shell vertex models more direct. 
} \la{move} \vspace{-.2cm} 
\end{figure}

Finally, the most important objects in (\ref{XXbar3pt}) are the hexagon vertices $h({\alpha|\gamma})$ and $h({\bar{\alpha}|\bar{\gamma}})$. We can relate either of them to a purely creation amplitude with all excitations in the top by applying a sequence of crossing transformations, see figure \ref{move}.
This leads to a representation of the hexagon as
\beq
h({\alpha|\beta})=\langle\mathfrak{h}| \prod_{u_j \in \alpha} X(u_j) \prod_{v_j \in \beta} \bar{X}(v_j^{-4\gamma}) \rangle\label{XXbarhex}
\eeq
According to our proposal (\ref{HexA}), the creation amplitude in (\ref{XXbarhex}) is given by a product of a dynamical and a matrix part, 
\beq
h({\alpha|\beta})= \texttt{dyn} \times \texttt{mat} \,.
\eeq
The dynamical part given by 
\beq
\texttt{dyn} = \frac{\prod\limits_{i<j} h(\alpha_i,\alpha_j) \prod\limits_{i<j} h(\beta_i,\beta_j)}{\prod\limits_{i,j} h(\beta_i,\alpha_j)}
\eeq
where we used the relations in appendix \ref{crossingFormulae} to cross the $v$-rapidities.

\begin{figure}[t]
\begin{center}
\includegraphics[scale=0.33]{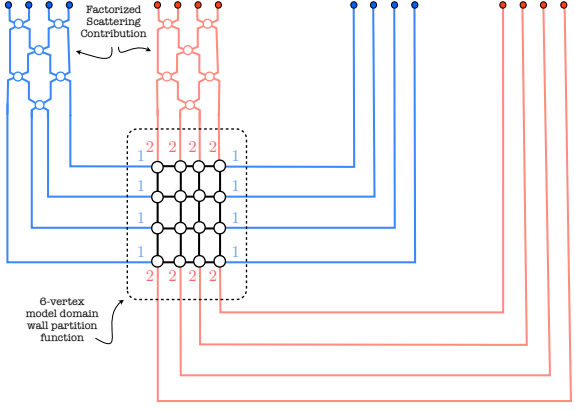}
\end{center}
 \vspace{-1.5cm} 
\caption{\normalsize The scattering between $\phi^1$'s (or between $\phi^2$'s) preserves the index structure and yields a totally factorized contribution. The most interesting contribution comes from the scattering between $\phi^1$s and $\phi^2$s, which, at weak coupling, can be identified with the domain wall partition function of the six-vertex model, see figure \ref{boltzmann}.
} \la{vertex} \vspace{-.2cm} 
\end{figure}

As also mentioned in that appendix, for the matrix part contribution we can ignore rotations by four mirror transformations (or two crossing transformations) which bring Zhukowsky variables back to themselves. As such, the matrix part contribution to (\ref{XXbarhex}) can be readily related to the scattering matrix element $\texttt{mat}=\<\texttt{out}| \mathcal{S}|\texttt{in}\rangle$
with 
\beqa
|\texttt{in}\rangle&=&|\phi^{1}(\alpha_1)\cdots\phi^{1}(\alpha_{|\alpha|})\phi^{2}(\beta_1)\cdots\phi^{2}(\beta_{|\beta|}) \rangle \nn \\
\langle\texttt{out}|&=&\langle \phi^{2}(\beta_{|\beta|})\cdots\phi^{2}(\beta_1) \phi^{1}(\alpha_{|\alpha|})\cdots \phi^{1}(\alpha_1) | \nn
\eeqa
%
and with the dressing phase set to one. This scattering process consists of two steps (see figure \ref{vertex}). The first step is the scattering among $\phi^1$s and the scattering among $\phi^2$s. Since the scattering of two $\phi^1$s (two $\phi^2$s) only produces $\phi^1$s ($\phi^2$s), this sub-process simply re-orders $\phi^1$s and $\phi^2$s and produces a phase factor given by a product of $h_{YY}$'s over $h_{DD}$'s. The second step is the scattering between $\phi^1$s and $\phi^2$s. At finite coupling, this is quite complicated since the scattering between $\phi^1$ and $\phi^2$ can produce fermions. However, since the production of fermions is suppressed by $O(g^2)$, we can ignore them in the weak-coupling limit. The matrix elements which survive in this limit then coincide with the Boltzmann weights of the six vertex model as depicted in figure \ref{boltzmann}. The resulting sub-process corresponds to the so-called domain wall partition function \cite{Ivan2, Izergin} which admits a well-known determinant representation \cite{Izergin}.

All in all, and combining with the (weak coupling leading order expansion of the) dynamical part we finally obtain the following expression for the hexagon process at tree level:
\beq
h(\alpha|\beta)=\frac{\det\displaystyle \left[ \frac{i}{(\alpha_n-\beta_m)(\alpha_n-\beta_m+i)}\right]}{\prod\limits_{i<j}(\alpha_i-\alpha_j)\prod\limits_{k>l}(\beta_k-\beta_l)/\prod\limits_{i,k }(\beta_k-\alpha_i-i)} \,.
\eeq
%
%
%
%
As a result, the structure constant (\ref{XXbar3pt}) is expressed as a sum over partitions with a summand given by a product of two determinants. This expression matches with the well known formula for off-shell scalar products \cite{korepin,Slavnov} and thereby reproduces the weak-coupling result.

It would be interesting to extend the analysis performed here to finite coupling and also to other sectors. For  general correlators, the vertex model of relevance is the full SU(2$|$2) vertex model, which coincides with the Hubbard model \cite{hubbard}. It seems very timely to revisit the computation of its partition functions. 
\begin{figure}[t]
\begin{center}
\includegraphics[height=6cm]{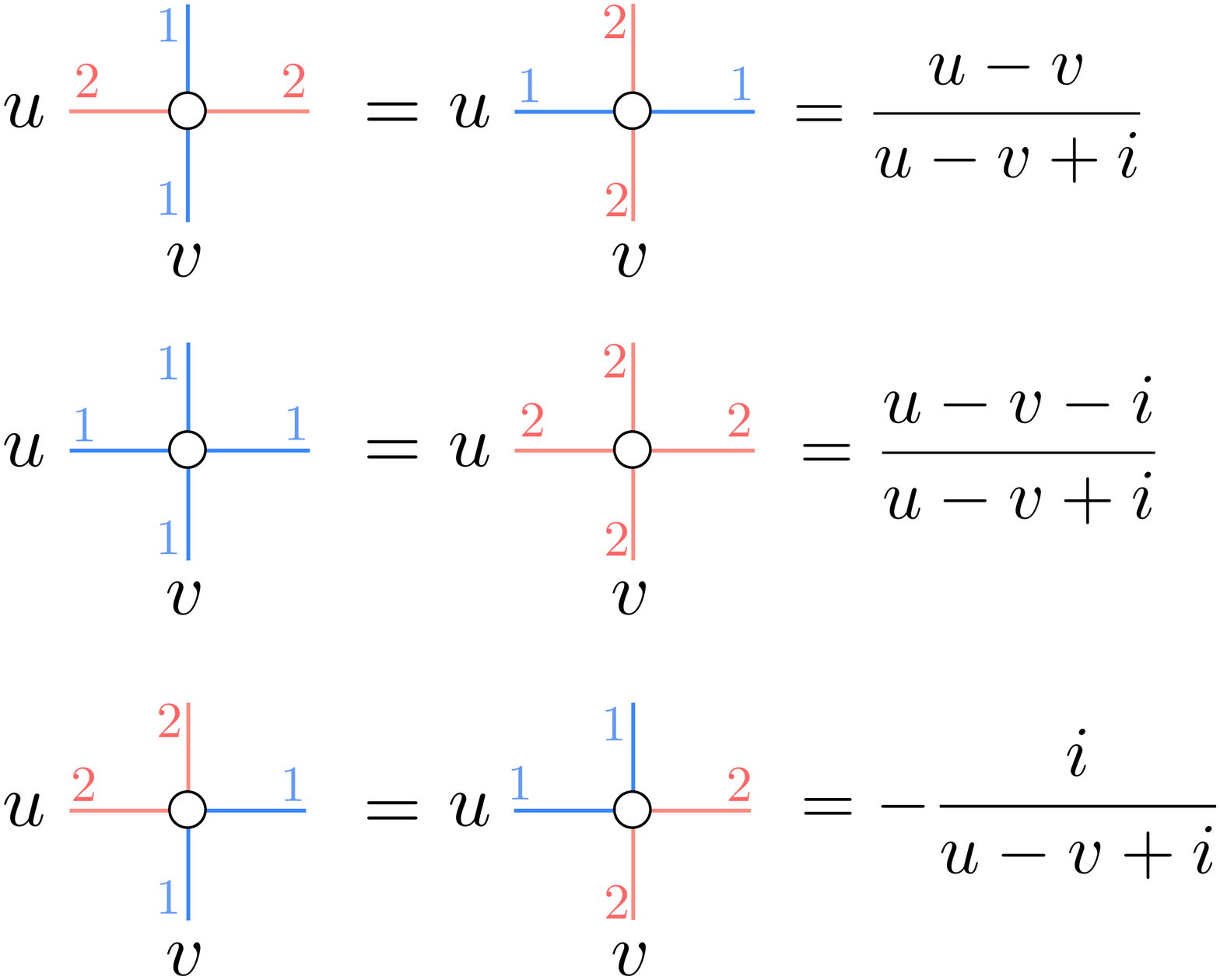}
\end{center}
\vspace{-.5cm}
\caption{\normalsize Matrix elements of the processes which are non-zero at weak coupling. They coincide with the Boltzmann weights of the rational six-vertex model. 
} \la{boltzmann} \vspace{-.2cm} 
\end{figure}

\section{Further nontrivial  $SU(2)$ checks}\label{Further}
Thus far, all our checks involved excitations on at most two of the three external operators. In this section we consider the most general case where all three operators are excited. A particularly symmetric configuration corresponds to inserting $Y$ scalars at each of the three operators. The corresponding set of Bethe roots are denoted as ${\bf u}_1,{\bf u}_2,{\bf u_3}$. This ought to be matched against the so-called correlators of type I-I-I studied in \cite{Kazama:2014sxa}. 

The result reported in \cite{Kazama:2014sxa}  for such correlators is quite rich. It involves a triple sum over partitions with a summand given by a product of three partial domain-wall partition functions, see formula (4.28) therein. In turn, each such partition function can be expressed by a sum of bipartite partitions. In the end, collecting together all relevant factors we quote here the result for such correlators:
\beqa
&&C_{123}^{\bullet\bullet\bullet} \propto\!\!   \sum_{\substack{\alpha_1\cup \bar{\alpha}_1 ={\bf u}_1\\ \alpha_2\cup \bar{\alpha}_2={\bf u}_2 \\ \alpha_3 \cup \bar{\alpha}_3={\bf u}_3}} 
\sum_{\substack{\beta_1\cup \bar{\beta}_1 = \bar{\alpha}_1 \cup \alpha_2 \\ \beta_2\cup \bar{\beta}_2 = \bar{\alpha}_2 \cup \alpha_3  \\ \beta_3\cup \bar{\beta}_3 = \bar{\alpha}_3 \cup \alpha_1 }} \mathbb{G}(\alpha_1,\dots,\bar{\beta}_3) \la{IIIres}
\eeqa
where
\beqa
&&\mathbb{G}= z_{12}^{N_3-|\bar\alpha_1|-|\alpha_2|}z_{23}^{N_1-|\bar\alpha_2|-|\alpha_3|}z_{31}^{N_2-|\bar\alpha_3|-|\alpha_1|}   \la{G}\\
&&\,\,\,(-1)^{|\alpha_1|+|\alpha_2|+|\alpha_3|+|\beta_1|+|\beta_2|+|\beta_3|}\frac{a_{\ell_{12}}(\alpha_2)a_{\ell_{23}}(\alpha_3)a_{\ell_{31}}(\alpha_1)}{a_{\ell_{12}}(\beta_1)a_{\ell_{23}}(\beta_2)a_{\ell_{31}}(\beta_3)}   \nn\\ 
&&\,\,\,f({\bar\alpha}_1,\alpha_1)f({\bar\alpha}_2,\alpha_2)f({\bar\alpha}_3,\alpha_3) f({\beta}_1,\bar\beta_1)f({\beta}_2,\bar\beta_2)f({\beta}_3,\bar\beta_3)\nn \,.
\eeqa
%
%
Here we are using the conventional notation where a function with a list as an argument stands for a product over the elements in the correspond lists. For instance
\beqa
f({\beta}_1,\bar\beta_1) = \prod_{z \in \beta_1} \prod_{w \in \bar\beta_1} f(z,w)  \,.
\eeqa
Finally $a_\ell(u) \equiv e^{i p(u) \ell}$ and $f(u,v) = 1+i/(u-v)$. Despite appearances, the above expression is independent of~$z_{ij}$.
Clearly, reproducing such involved result would constitute quite a nontrivial check of the hexagon approach! 

 \begin{figure}[t]
\begin{center}
\includegraphics[scale=0.31]{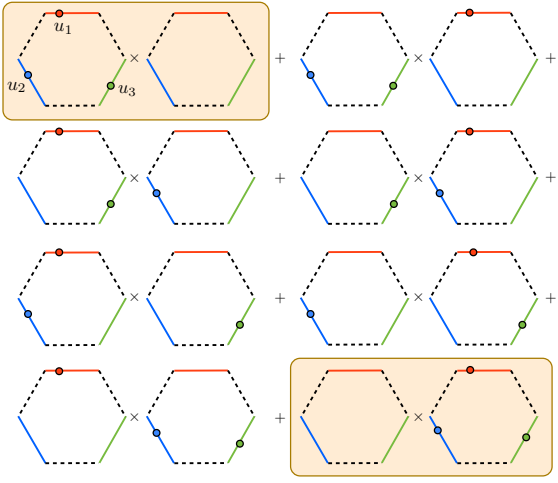}
\end{center}
\vspace{-0.5cm}
\caption{\normalsize Each of the three operator has a single excitation which can end up on the left or right hexagons upon cutting the pair of pants. Each such possibility can be matched to each of the eight terms in (\ref{goal}) when ${\bf u}=\{u_1\}$, ${\bf v}=\{u_2\}$ and~${\bf w}=\{u_3\}$. 
} \la{eightterms} \vspace{-.2cm} 
\end{figure}

The way this check goes requires some further simple massaging of (\ref{G}). More precisely we should slightly re-write the $a$-factors in the denominator in (\ref{G}). Take for instance $a_{\ell_{12}}(\beta_1)$. Since $\beta_1\cup \bar{\beta}_1 = \bar{\alpha}_1 \cup \alpha_2 \subset {\bf u}_1 \cup {\bf u}_2$, this factor is a product of $a_{\ell_{12}}$'s with some arguments belonging to ${\bf u}_1 $ and some belonging to ${\bf u}_2$. The latter will cancel some terms in the numerator factor $a_{\ell_{12}}(\alpha_2)$ in (\ref{G}) while the latter is eliminated in favour of $a_{\ell_{31}}(u_{1,j})$ by using Bethe equations
\beq
a_{\ell_{12}}(u_{1,j})a_{\ell_{31}}(u_{1,j}) = \prod_{i \neq j } \frac{f(u_{1,j},u_{1,i})}{f(u_{1,i},u_{1,j})} 
\eeq
The other two factors in the denominator are massaged in the same way. Note that in doing this we managed to bring all $a_\ell$ factors to the numerator and furthermore, $a_{\ell_{ii+1}}$ always takes as arguments roots of the the $i+1$-th operator. More precisely, what we observed by inspection on a few simple examples is that we can, in the end, 
recollect (\ref{IIIres}) into a single sum of the form  
\beqa
 C_{123}^{\bullet\bullet\bullet} \propto\!\!\!   \sum_{\substack{\delta_1\cup \bar{\delta}_1 ={\bf u}\\ \delta_2\cup \bar{\delta}_2={\bf v} \\ \delta_3 \cup \bar{\delta}_3={\bf w}}} \!\! 
(-1)^{\sum\limits_{i=1}^3\! |\bar \delta_i|} \, w_{\ell_{12}}(\delta_2,\bar \delta_2) w_{\ell_{23}}(\delta_3,\bar \delta_3) w_{\ell_{31}}(\delta_1,\bar \delta_1)  \nn \\ \!\!\!\!\!\! \!\!\!\!\!\! \!\!\!\!\!\! \!\!\!\!\!\! \!\!\!\!\!\! \times\,\, H(\delta_1|\delta_2|\delta_3)H(\bar\delta_1|\bar\delta_2|\bar\delta_3)\,,\qquad\qquad \la{goal}
\eeqa
with 
\beq
w_\ell(\alpha,\bar \alpha) = \prod_{ u_j \in \bar \alpha} (a_\ell(u_j)  \prod_{u_i \in \alpha \text{ with }  i > j}  S(u_j,u_i) ) \,,
\eeq
the splitting factor -- see figure \ref{Example2}. Once brought to the form (\ref{goal}) we can directly compare $H(\delta_1|\delta_2|\delta_3)$ with our prediction for an hexagon process with particles with rapidities $\delta_i$ at the i-th physical edge (and similar for the second hexagon).

%
%
For illustration, consider the simplest of such comparison for a single excitation on each operator~\cite{strictly}. In this case we have $2^3=8$ terms, see figure \ref{eightterms}. Of those, six correspond to having one hexagon with one excitation and another hexagon with two excitations. These should match with the same sort of hexagon processes already probed in section \ref{simpleWeak} and indeed they do. (Conversely, it is also useful to use the knowledge of these terms to fix the normalization.) The remaining two terms, highlighted in figure \ref{eightterms}, are the most interesting ones: they allow us to read off the three particle hexagon process with one particle in each physical edge. We observe that it is no longer of a simple factorized form. Instead it reads
\beq
H(u,v,w)= 1+\frac{i}{u-v}+\frac{i}{v-w}+\frac{i}{w-u} \,. \la{111gaugeRes}
\eeq
This ends the data analysis. We now turn to the hexagon side and try to recover this result. 

To compute such hexagon process we can start with three scalars at the top edge and move two of them down using mirror transformations. As explained in appendix \ref{mapping}, such crossing manipulations of scalars often entail dressing the naive expression by a few momentum factors. In the case at hand, for instance, we have (see figure \ref{monodromies2} for a very similar manipulation) 
\beqa
&&H(u,v,w)\equiv \langle\mathfrak{h}|\Big(|Y(u) \rangle \otimes |Y(v)\>\otimes |Y(w)\>\Big)\\
&&\,\,\,\, =-e^{i p(w)-i p(u)}\times \langle\mathfrak{h}|\Big( |Y(w^{4\gamma})\bar{Y}(v^{2\gamma})Y(u)\rangle\otimes |0\>\otimes |0\> \Big)\,. \nn
\eeqa
According to our ansatz (\ref{HexA}), the second factor in the second line can be trivially computed as
\beqa
&&\langle\mathfrak{h}|\Big( |Y(w)\bar{Y}(v)Y(u)\rangle\otimes |0\>\otimes |0\> \Big) =  \la{voilaSP} \\
&&\qquad h_{DD}(w,v)h_{DD}(w,u)h_{DD}(v,u) \times  \nn \\ 
&&\qquad\qquad \<\phi^{2}(u)\phi^{1}(v)\phi^2(w)| \mathcal{S}|\phi^{1}(w)\phi^{2}(v)\phi^1(u) \nn\rangle 
\eeqa
%
%
The matrix element can be straightforwardly evaluated. We get
\beqa
&&\!\!\!\!\!\!\!\!\!\!\!\!\!\!\!\! 8\<\phi^{2}(u_3)\phi^{1}(u_2)\phi^2(u_1)| \mathcal{S}|\phi^{1}(u_1)\phi^{2}(u_2)\phi^1(u_3) \rangle = \nn \\
&&  B_{12} \left(B_{13} \left(A_{23}+B_{23}\right)-A_{13} \left(A_{23}-3 B_{23}\right)\right) \nn \\
&+&A_{12}
   \left(A_{13} \left(3 A_{23}-B_{23}\right)+B_{13} \left(A_{23}+B_{23}\right)\right) \nn \\
&+&   4\,C_{12} K_{13} F_{23}  \times \frac{
x^-_1 x_3^+}{x_1^+ x_3^- } \la{finaluff}
\eeqa
where the two middle lines come from scattering of bosons while the last line comes from a fermion loop, see figure \ref{111-fig}. 

It remains to cross $w$ in (\ref{voilaSP}) twice and $v$ once which we can straightforwardly do, see appendix \ref{crossingFormulae}. Finally, we should expand the result at weak coupling. When the dust settles we perfectly reproduce (\ref{111gaugeRes}). (In particular it is curious to note that the fermion loop does contribute due to the non-commutativity of the weak coupling expansion with mirror transformations.)
%
%
As mentioned above, we repeated this sort of analysis for more complicated cases, involving more excitations. As should be clear by now, the corresponding analysis is more cumbersome but equally straightforward. 
\begin{figure}[t]
\begin{center}
\includegraphics[height=2.8cm]{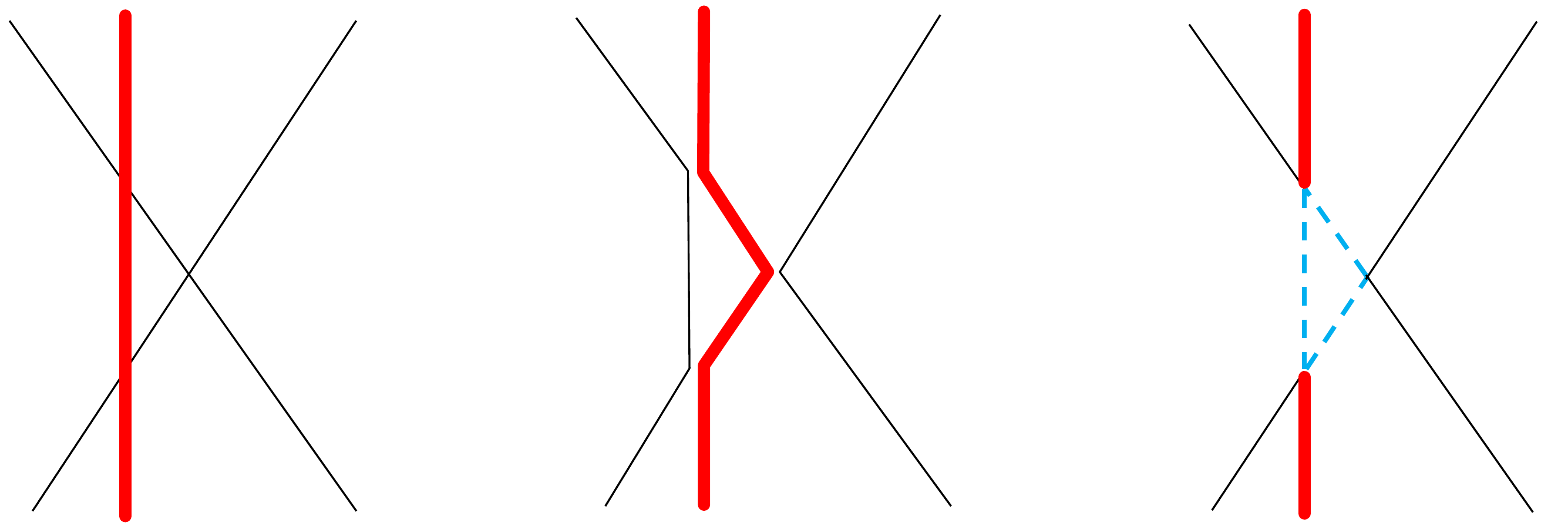}
\end{center}
\caption{\normalsize Three different processes contributing to the S-matrix element  (\ref{finaluff}). The black lines denote bosons with the index $1$ and the thick red lines denote bosons with the index $2$, whereas the blue dashed lines denote fermions.
} \la{111-fig} \vspace{-.2cm} 
\end{figure}

Intriguingly, the computation based on the hexagon appears to be quite different from the gauge-theory one: On the gauge-theory side, we only need to use the bosonic $SU(2)$ spin chain whereas in the hexagon approach, the effect of fermions is essential to obtain the correct result. It would be interesting to further explore and clarify the relation between two approaches.

It would be very interesting to reproduce (\ref{IIIres}) -- and other such involved tree level correlators available in the literature (see e.g. \cite{tailoring1,Ivan2,tailoringNC,ThiagoJoao,su3,Kazama:2014sxa,volodya}) --  in full generality from the hexagon approach.

\section{Wrapping effects from the classical string} \la{strongAp}

The structure constants at strong coupling were computed in \cite{Janik:2011bd,shotasu2,Kazama:2013qsa,shota2} using the classical string assuming the existence of a smooth saddle-point solution. The result obtained there is given by integrals similar to the ones at weak coupling if not for the contours. 
(These contours are amongst the most subtle points of the strong coupling analysis and were properly cleaned up very recently \cite{shota2}.)
As explained in \cite{shota2}, the strong coupling contours consist of two parts: the ones around the branch cuts formed by the condensation of the Bethe roots and the ones around the unit circle. As discussed in the main text, the ones around the Bethe roots reproduces the asymptotic part of the hexagons. In this appendix, we will see that the ones around the unit circle reproduces the finite-size correction along the bridges.

The part relevant for the comparison is given by
\beq
\oint_{U} \frac{du}{2\pi} \left[{\rm Li}_{2} \left( e^{i(\hat p_2+\hat p_3-\hat p_1)}\right)-{\rm Li}_{2} \left( e^{i(\tilde p_2+\tilde p_3-\tilde p_1)}\right)\right]\label{li2fin}
\eeq
where $\hat p_{i}$ is the AdS-part of the quasi-momentum and $\tilde p_{i}$ is the S$^5$-part of the quasi-momentum. The contour of integration $U$ encircles the unit circle in the $x$ plane counterclockwise. 

As in the main text, here we are considering the situation where only $\mathcal{O}_1$ is non-BPS and contains excitations in the AdS-part such that $\tilde p_1=L_1 x/(2g(x^2-1))$. The other two operators are BPS and thus their sphere and AdS quasi-momenta are identical, $p_2=\hat p_2 = \tilde p_2 = L_2 x/(2g(x^2-1))$ and similar for $p_3$. 

When the size of the bridge $\ell_{23}$ is large, the argument of the dilogarithm is exponentially small on the lower semi-circle and can be naturally expanded using
\beq
{\rm Li}_2 (z) \sim z +\cdots \quad z\sim 0.
\eeq
On the other hand, the argument blows up on the upper semi-circle and produces  divergent contributions. 
However, those terms eventually cancel out if we sum up all the terms contributing to the structure constant. Therefore, we can safely remove such pieces and focus on the remaining exponentially small parts. This corresponds to  expanding the dilogarithm at infinity as
\beq
{\rm Li}_2 (z)\sim -\frac{1}{2}\log^2 (-z)-\frac{\pi^2}{6}-z^{-1}+\cdots \quad z\sim \infty
\eeq
and discarding the first two terms.
After this manipulation, the equation (\ref{li2fin}) yields the following leading order correction:
\beqa
&&\oint_{U^{-}}\frac{du}{2\pi} \left(e^{i(\hat p_2+\hat p_3-\hat p_1)}-e^{i(\tilde p_2+\tilde p_3-\tilde p_1)}\right)\nn \\
&&\qquad -\oint_{U^{+}}\frac{du}{2\pi} \left(e^{-i(\hat p_2+\hat p_3-\hat p_1)}-e^{-i(\tilde p_2+\tilde p_3-\tilde p_1)}\right)\label{notnice2}
\eeqa
Here $U^{+}$ and $U^{-}$ denote the upper and lower semi-circle respectively. By changing the integration variable for the second term from~$x$ to~$1/x$ and using~$\tilde p_i(1/x)=-\tilde p_i(x)$ under this transformation, we can re-write (\ref{notnice2}) as in~(\ref{mirror-strong}) in perfect agreement with the hexagon approach.

\section{Impossibilities}\la{Impossible}

In (appendix D.2 of) \cite{tailoring4} it was conjectured that 
\beq
\mathcal{A}(u_1,\dots,u_{\ell+1})=\mathcal{O}(g^{2\ell+2}) \la{Aconjecture}
\eeq
in the $SU(2)$ case. The motivation for this conjecture was very simple: it is obviously impossible to pass $\ell+1$ scalars through a bridge of size~$\ell$. In \cite{tailoring4} it was argued that if an all-loop generalization of $\mathcal{A}$ exists it will probably hold in an asymptotic sense. As such it was proposed that such contribution should vanish only up to wrapping corrections. Hence the right hand side in~(\ref{Aconjecture}). The hexagon conjecture consolidates nicely such ideas. Indeed, our proposal is that $\mathcal{A}$ describes the all-loop asymptotic part of the result. With an all-loop proposal at hand, we checked (\ref{Aconjecture}) to very high loop order (four loops). This depends crucially on the form of $h(u,v)$ and is thus a very non-trivial self-consistency check of our proposal at higher loops which does not require explicit comparison with data. Along the same lines we checked a few other impossibilities. We considered, for instance, a correlator with $\ell+1$ scalars $X$ and the bottom operator $\mathcal{O}_2$ and the same number at the top operator $\mathcal{O}_1$. Those scalars need to talk to each other through the bridge of size $\ell$ which should again be impossible. To check this is more complicated since it now entails computing two sums over partitions (one for the top and one for the bottom). Nicely, the conclusion is the same as before: Such configurations yield zero up to wrapping order. We repeated the same analysis with these scalars replaced by fermions. Again, by the very same logic, we expect to get non-trivial cancelations yielding a zero result up to wrapping correction for this case as well. This worked perfectly once again up to the order we checked (four loops). For derivatives there is no such obvious constraint since derivatives can pile up and pass through any bridge of any size. It would be interesting to investigate whether there is still some impossible Gedanken experiment one could imagine in that case. It would also be interesting to investigate if by including wrapping corrections we could extend such ideas further and obtain further constraints/checks of the hexagon construction. 

\section{pp-wave SFT vertex as a twist operator\label{ap-b}}
Here we show that the pp-wave string field theory (SFT) vertex, constructed in \cite{SFTvertex} and analyzed in \cite{Bajnok:2015hla} from the integrability viewpoint, can be understood as the form factor of the octagon twist operator of the free massive boson.

The primary focus of the analysis in \cite{Bajnok:2015hla} was on the cases where the outgoing state (string $3$)  and one of the incoming states (string $2$) are infinitely long whereas the other incoming state (string $1$) has a finite size. However, to apply the integrable bootstrap program, it is more convenient to consider the situations where all three strings are infinitely long. The Neumann coefficients in this limit are also computed in \cite{Bajnok:2015hla} and, up to normalization, they read
\beq
\begin{aligned}
N^{33}(\theta,\theta')&\propto \frac{(1-e^{ipL})(1-e^{ip^{\prime}L})}{\cosh  \frac{\theta-\theta'}{2}},\\
N^{32}(\theta,\theta')&\propto i\frac{(1-e^{ipL})}{\sinh \frac{\theta-\theta'}{2}},\\
N^{22}(\theta,\theta')&\propto \frac{1}{\cosh \frac{\theta-\theta'}{2}}.
\end{aligned}\label{Neumann}
\eeq

There are two intriguing features in these expressions: One is the universal factor $\cosh (\theta-\theta')/2$ (or $\sinh (\theta-\theta')/2$) in the denominator and the other is the factor in the numerator $(1-e^{ipL})$, which appears when the excitation is on the outgoing string. The universal factor can be understood as the two-particle form factor of the branch-point twist operator in the free massive boson, which was computed in \cite{Cardy:2007mb}. When the excess angle is $2\pi (n-1)$, the two-particle creation form factor takes the following form:
 \beq
F_2 (\theta-\theta^{\prime})\propto \frac{1}{\sinh \frac{i\pi +\theta-\theta'}{2n} \sinh \frac{i\pi -\theta+\theta'}{2n}}\label{twistFF}
 \eeq
 Upon setting $n=2$, as needed for an octagon, this expression reproduces the denominator in (\ref{Neumann}). (Note that implicit here is the assumption that the minimal octagon form factor is an appropriate choice, which requires in particular that the associated octagon vertex preserves the flavour symmetries of the theory. For comparison, the above form factor with $n=3/2$ would not be a good choice for our hexagon at strong coupling, since it does not incorporate the breaking of the symmetry group down to the diagonal subgroup.)

On the other hand, the numerator factor $(1-e^{i p L})$ arises from the sum over partitions as we see below. When we take the decompactification limit of the lightcone diagram, we need to cut two edges of the worldsheet and open it up into the octagon. Upon doing so, the outgoing string is cut into two halves whereas the two incoming strings are simply cut open into open segments. As in the case of the hexagons discussed in the main text,  the excitations on the outgoing string can end on either half and we need to sum over all possibilities taking into account the extra factor coming from the phase shift. When the outgoing string has one excitation, this yields $(1-e^{ipL})$ reproducing the factor present in $N^{32}$. On the other hand, when there are two excitations, the sum over partition leads
\beq
1-e^{ipL}-e^{ip^{\prime}L}+e^{i(pL+p^{\prime}L)},
   \eeq
   which coincides with the numerator of $N^{33}$.
   
Before ending this appendix, let us make a few comments. When the strings have finite lengths, the Neumann coefficients are corrected by terms of order $e^{-m L_i}$. These corrections should result from the exchange of excitations between the mirror channels of the octagon. It would be interesting to study in detail how the integration over the mirror excitations reproduces the exact finite-size results given in~\cite{SFTvertex}. Also, the connection between the string interaction and the twist operators was explored previously in the context of matrix string theory \cite{MatrixString}. Understanding the relation with the approach discussed in this paper would be interesting. 




\begin{thebibliography}{99}
\bibitem{stat}
Explicitely $\frak{f} = \sum_{i<j}\dot{f}_{i}f_{j}$ where $f_{i}$ ($\dot f_{i}$) is the fermion number of $i$-th left (right) component.
\bibitem{norm}

The overall normalization in~(\ref{1-point}) is of no concern here, since a properly normalized three-point function is insensitive to it. More technically, its effect on general form factors is multiplicative and should be exactly compensated by the effect of multiplying with the measures $\mu$ as in~(\ref{rank1}).
\bibitem{trivial} Note that we include here the trivial (graded) scattering $(-1)^{F/\dot{F}}$ between left with right movers.
\bibitem{comment} Note that solutions always come in pair since if $h_{12}$ is a solution so is $h'_{12}=1/h_{21}$. The solution we picked is the regular one in the limit $2\rightarrow 1$.
\bibitem{alternativesign}
  There is an alternative way to eliminate the minus signs, which is to identify the states as $\phi^{2\gamma}=\phi^{-2\gamma}$ and $\psi^{2\gamma}=-\psi^{-2\gamma}$. However, a notable difference is that, while (\ref{2gamma-2}) preserves the one-particle hexagon, the one above does not. It would be important to understand why (\ref{2gamma-2}) has nicer properties.
\bibitem{strictly} Strictly speaking such configuration is illegal by the zero momentum condition on each operator; here we proceed as in section \ref{simpleWeak} and relax  zero momentum  while still imposing Bethe equations.
\end{thebibliography}
\end{document}